\documentclass[journal,draftclsnofoot,onecolumn,12pt,twoside]{IEEEtran}
\usepackage{mathrsfs}
\usepackage{amsfonts}
\usepackage{graphicx,cite,epsfig,amssymb,amsmath}
\usepackage{color,xcolor}
\usepackage{pifont}
\usepackage{stmaryrd}
\usepackage{setspace}
\usepackage{subfigure}
\usepackage{cite}
\usepackage{array}
\usepackage{float}
\usepackage{multirow}
\usepackage[ruled,linesnumbered]{algorithm2e}
\usepackage{epstopdf}
\usepackage{mathtools}
\usepackage{multirow}
\usepackage{framed}
\usepackage{ntheorem}
\theoremseparator{.}
\theorembodyfont{\upshape}
\floatstyle{ruled}

\providecommand{\algorithmname}{Algorithm}
\floatname{algorithm}{\protect\algorithmname}

\newtheorem{thm}{\bf Theorem}

\newtheorem{lem}{\bf Lemma}

\newtheorem{coro}{\bf Corollary}
\newtheorem{defi}{\bf Definition}
\newtheorem{assu}{\bf Assumption}
\def\proof{{\emph{Proof:} }}
\begin{document}
\title{Scheduling for Cellular Federated Edge Learning with Importance and Channel Awareness}
\author{Jinke Ren, Yinghui He, Dingzhu Wen, Guanding Yu,\\ Kaibin Huang, and Dongning Guo
	\thanks{J. Ren is with the College of Information Science and Electronic Engineering, Zhejiang University, Hangzhou 310027, China, and also with the Department of Electrical and Computer Engineering, Northwestern University, Evanston, IL 60208 USA (e-mail: renjinke@zju.edu.cn).}
    \thanks{Y. He and G. Yu are with the College of Information Science and Electronic Engineering, Zhejiang University, Hangzhou 310027, China (e-mail: 2014hyh@zju.edu.cn, yuguanding@zju.edu.cn).}
	\thanks{D. Wen and K. Huang are with the Department of Electrical and Electronic Engineering at The University of Hong Kong, Hong Kong (e-mail: dzwen@eee.hku.hk, huangkb@eee.hku.hk).}
    \thanks{D. Guo is with the Department of Electrical and Computer Engineering, Northwestern University, Evanston, IL 60208 USA (e-mail: dGuo@northwestern.edu).}
    \thanks{This material is based upon work supported by the National Science Foundation under Grants No. CCF-1910168 and CNS-2003098.}}%\emph{Corresponding author: XXX.}
	 %\IEEEaftertitletext{\vspace{-0.75\baselineskip}}
\maketitle
%\vspace{-3em}
\vspace{-3.6em}
\begin{abstract}
In cellular federated edge learning (FEEL), multiple edge devices holding local data jointly train a neural network by communicating learning updates with an access point without exchanging their data samples. With very limited communication resources, it is beneficial to schedule the most informative local learning updates. In this paper, a novel scheduling policy is proposed to exploit both diversity in multiuser channels and diversity in the ``importance" of the edge devices' learning updates. First, a new probabilistic scheduling framework is developed to yield unbiased update aggregation in FEEL. The importance of a local learning update is measured by its gradient divergence. If one edge device is scheduled in each communication round, the scheduling policy is derived in closed form to achieve the optimal trade-off between channel quality and update importance. The probabilistic scheduling framework is then extended to allow scheduling multiple edge devices in each communication round. Numerical results obtained using popular models and learning datasets demonstrate that the proposed scheduling policy can achieve faster model convergence and higher learning accuracy than conventional scheduling policies that only exploit a single type of diversity.
\end{abstract}
\vspace{-0.3em}
\section{Introduction}
As driven by the phenomenal growth in global mobile data traffic, \textit{artifical intelligence} (AI) has revolutionized almost every branch of science and technology ranging from computer vision, natural language processing, to wireless networks \cite{AI_in_Wireless_Survey_1,6G_Chen, AI_in_Wireless_Survey_2, AI_in_Wireless_Survey_3}. The implementation of AI in wireless networks is foreseen as an innovative breakthrough for realizing an intelligent network. To enable rapid access to the large amount of real-time data generated by massive distributed edge devices, a new paradigm of computing, called \textit{edge learning} has emerged by migrating learning from central clouds towards the edge of wireless networks \cite{Edge_Learning, Network_Intelligence_at_the_Edge, Edge_AI_ChenXu}. Despite the proximity to data source in edge learning, it still faces with a critical challenge of privacy concern when transmitting raw data to the edge server. To overcome this deficiency, an innovative framework namely \textit{federated edge learning} (FEEL) has been proposed, which features distributed learning at edge devices and update aggregation at the edge server \cite{Communication_Efficient_from_Decentralized_Data}. By periodically reporting the local learning updates to the edge server for global aggregation, FEEL is capable of leveraging the heterogeneous data to attain an accurate learning model.

A major design objective in FEEL is to accelerate the training process and maintain certain learning accuracy under communication and computation resource constraints. Due to that communication suffers from impairment of wireless channels (e.g., channel fading, interference, and noise), it is often the main bottleneck for fast FEEL \cite{Client_Selection, My_Work,Hierarchical_FL, Added_2}. This fact calls for effective scheduling algorithms for \textit{fast update acquisition} from highly distributed edge devices. Conventional scheduling principles focus on data rate \cite{Data_Rate} or quality-of-service \cite{QoS} whereas assume that the transmitted data is equally important. However, this assumption is often inappropriate in FEEL since different local learning updates are of dissimilar significance to the model convergence \cite{Importance_RRM, Lag,Added_3}. This motivates researchers to design a new scheduling approach based on not only the channel states but also how important the local updates are for learning. Towards this end, we consider the scheduling problem in FEEL by jointly exploiting the diversity involved in both wireless channels and local learning updates.

\subsection{Prior Work}
Communication-efficient FEEL requires the joint design of learning algorithms and communication techniques. Recent years have seen much research interests from both industry and academia, covering key topics such as learning update compression \cite{Two_Strategies,Gradient_Compression_Survey,Deep_Gradient_Compression}, radio resource management \cite{Chen_Packet_Error,Local_Global_Tradeoff,Scheduling_Bennis,Energy_Delay_Tradeoff_Inforcom,Energy_Delay_Tradeoff_arxiv}, and over-the-air computation \cite{AirComp_Ding,BAA,Zhu}. Specifically, the authors of \cite{Two_Strategies} proposed two update compression approaches based on random sparsification and probabilistic quantization to reduce the communication cost between edge server and edge devices. In \cite{Gradient_Compression_Survey}, a sparse ternary compression framework was developed to adapt to the non-independent and identically distributed (non-IID) FEEL environment, which compresses the upstream and downstream via sparsification, ternarization, error accumulation, and Golomb encoding. Moreover, the deep gradient compression technology was employed in \cite{Deep_Gradient_Compression}, where a hierarchical threshold selection policy was proposed to speed up the gradient sparsification process. On the other hand, the authors of \cite{Local_Global_Tradeoff} developed an efficient control algorithm to achieve the trade-off between local update and global aggregation during the learning process with limited communication resource budget. Further, a joint user selection and resource allocation policy towards minimizing the FEEL loss function was proposed in \cite{Chen_Packet_Error} by taking into account the packet error over wireless links.
In addition, \cite{Scheduling_Bennis} presented an effective client scheduling and resource allocation policy for FEEL over wireless links with imperfect channel state information (CSI), and \cite{Scheduling_Chen} proposed a joint user selection and resource allocation policy to minimize the FEEL convergence time based on probabilistic scheduling.
%In addition, \cite{Scheduling_Bennis} presented an effective client scheduling and resource allocation policy for FEEL over wireless links with imperfect channel state information (CSI).
Besides, the trade-off between learning time and energy consumption in FEEL was investigated in \cite{Energy_Delay_Tradeoff_Inforcom} and \cite{Energy_Delay_Tradeoff_arxiv}, where \cite{Energy_Delay_Tradeoff_Inforcom} developed a closed-form communication and computation resource allocation in a synchronous manner, and \cite{Energy_Delay_Tradeoff_arxiv} designed an iterative algorithm for joint power control and resource allocation. In particular, a novel technique namely \textit{over-the-air computation} is adopted for update aggregation in FEEL. Based on this, the authors of \cite{AirComp_Ding} considered joint device selection and beamforming design in FEEL. Moreover, a prominent broadband analog aggregation multiple-access scheme featuring the aggregation of analog modulated gradients was proposed in \cite{BAA}, where two communication-and-learning trade-offs were highlighted as well. To facilitate implementation,\cite{BAA} was further extended in \cite{Zhu}, which designed a digital broadband over-the-air aggregation scheme to evaluate the effect of wireless channel impairment on model convergence.

On the other hand, the local learning updates in FEEL are not equally important for model convergence and thus are valuable to be exploited for further performance improvement \cite{ZhangTong}. Some prior works incorporated the new feature of data importance for communication design in centralized edge learning systems, such as data retransmission \cite{Retransmission} and user scheduling \cite{Scheduling_Liu}. Specifically, a pioneering importance-aware automatic-repeat-request protocol was designed to balance the communication reliability and the data uncertainty \cite{Retransmission}. Furthermore, the authors also proposed an importance-aware user scheduling scheme to improve the communication efficiency in edge learning \cite{Scheduling_Liu}. Inspired by this, we consider to exploit the update importance for scheduling design in FEEL, which has not been well studied in existing works.
\subsection{Contributions and Organization}
In this paper, we propose a novel importance- and channel-aware scheduling policy to improve the performance of FEEL. This work is most related to the prominent FEEL works of \cite{Scheduling_Tony} in analyzing convergence behaviour using classical scheduling policies and \cite{Scheduling_Poor} in combining channel state and update norm for scheduling. However, the scheduling policy in \cite{Scheduling_Tony} is not specifically designed for FEEL while the convergence analysis is not provided in \cite{Scheduling_Poor}. We go one further step than \cite{Scheduling_Tony} and \cite{Scheduling_Poor} in that we provide fundamental convergence analysis as well as develop an innovative scheduling for FEEL.
%This work is most related to the prominent FEEL works of \cite{Scheduling_Tony} in analyzing convergence behaviour using classical scheduling policies and \cite{Scheduling_Poor} in combining channel state and update norm for scheduling. We go one further step than \cite{Scheduling_Tony} and \cite{Scheduling_Poor} in that we provide fundamental analysis for innovative scheduling design.
Our main result demonstrates that the scheduling decision should elegantly balance the diversity in wireless channels and learning updates to achieve a satisfactory learning performance.

The main contributions of this work are summarized as follows.
\begin{itemize}
  \item We first propose a novel probabilistic scheduling framework to yield unbiased gradient aggregation in FEEL. The one-round convergence rate is analyzed and the update importance is measured by the gradient divergence.
  \item We consider a simple strategy of scheduling one device in each communication round and devise an importance- and channel-aware scheduling policy. The corresponding convergence rate is also provided. More precisely, the scheduling probability increases linearly with both the data unbalanced indicator and the local gradient norm, whereas decreases sublinearly with the exponent of $-\frac{1}{2}$ when the local gradient upload latency is large.
  \item We further extend the probabilistic scheduling framework to the strategy of scheduling multiple devices in each communication round. An efficient scheduling without replacement algorithm is designed, followed by the closed-form expression of bandwidth allocation.
  %\item We evaluate the performance of the proposed scheduling policy via extensive experiments using some popular datasets.
  Numerical results demonstrate gains achieved by the proposed scheduling scheme over conventional scheduling schemes.
\end{itemize}

The rest of the paper is organized as follows. Section II introduces the system model and motivates the learning mechanism. Section III investigates the simple strategy of scheduling one device in each communication round and formulates the probabilistic scheduling problem. The optimal solution and the convergence analysis are provided in Section IV. Section V discusses how to extend the probabilistic scheduling to the strategy of scheduling multiple devices in each communication round. Section VI draws the numerical results, followed by conclusions in Section VII.
\section{System Model and Learning Mechanism}
\subsection{Federated Edge Learning System Model}
\begin{figure}[htp]
	\centering
	\includegraphics[width=6.2in]{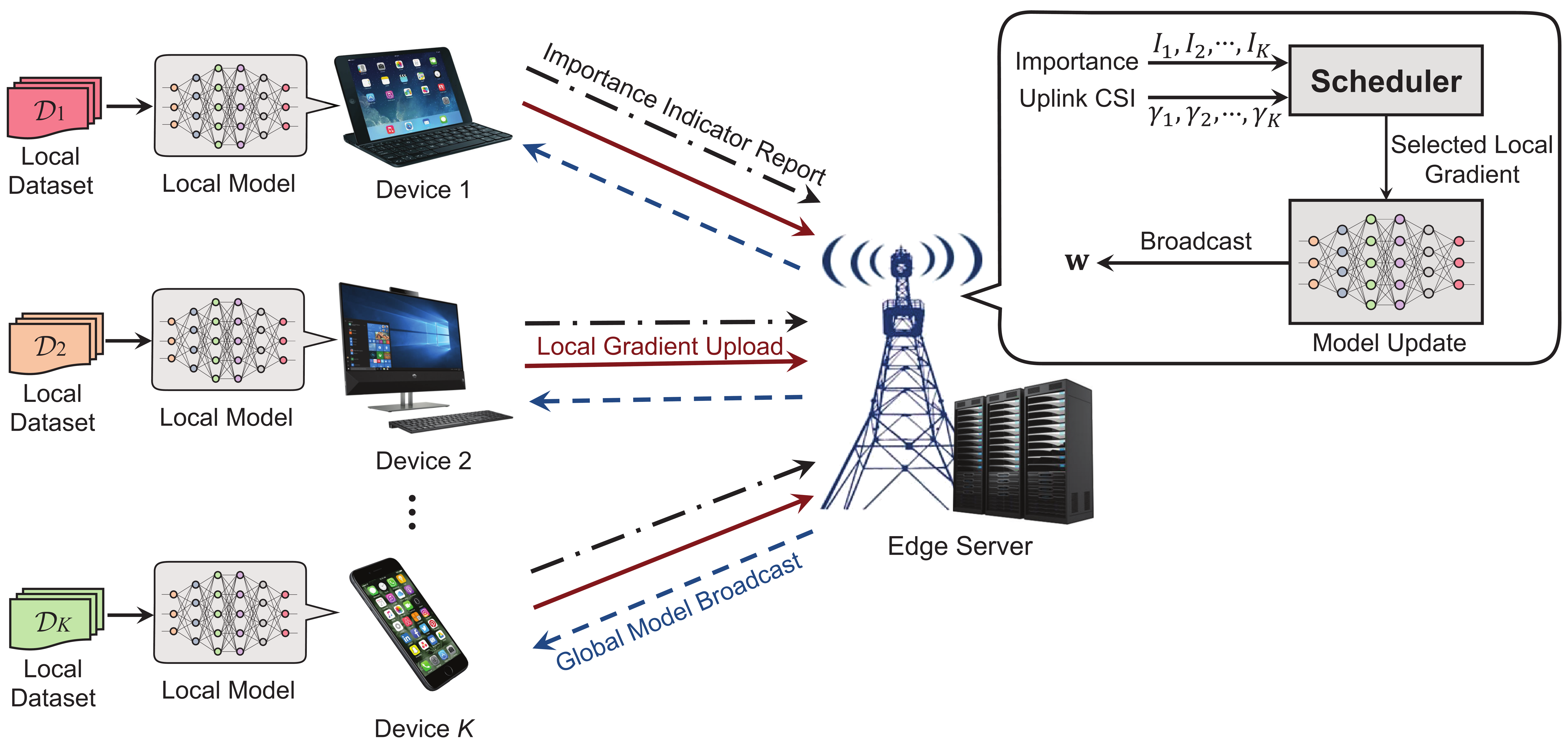}
	\caption{Federated edge learning system.}	
	\label{FEL system}
\end{figure}
As illustrated in Fig. \ref{FEL system}, we consider an FEEL system comprising one edge server and $K$ decentralized single-antenna edge devices, denoted by a set $\mathcal{K}=\left\{1, 2, \dots, K\right\}$. Each device has a fraction of labeled data. Let $\mathcal{D}_k = \left(\left({\bf{x}}_{k}^1,y_{k}^1\right),\left({\bf{x}}_{k}^2,y_{k}^2\right),\cdots,\left({\bf{x}}_{k}^{n_k},y_k^{n_k}\right)\right)$ denote the local dataset of device $k$, where ${\bf{x}}_k^i$ is its $i$-th training data sample, $y_k^i$ is the corresponding ground-truth label, and $n_k$ is the number of data that device $k$ owns. In view of the heterogeneous data structure, we assume that local datasets are statistically independent across edge devices.
%In view of the heterogeneous data structure, we assume that the local datasets are independent of each other.
Let $n=\sum_{k=1}^K n_k$.  A shared learning model, denoted by $\bf{w}$, needs to be collaboratively trained across all edge devices with their distributed local datasets. To leverage the rich data over edge devices while preserving data privacy, all devices adopt the federated learning technique by iteratively reporting their locally computed gradient for global aggregation in lieu of raw data. Due to the gradient sparsity, the communication overhead can be significantly reduced by some gradient compression approaches. Therefore, we focus on the gradient-averaging implementation in the subsequent exposition.
%\footnote{In this paper, we focus on the gradient-averaging method while both the design principle and the analytical results can be extended to the alternative model-averaging method.}
 Moreover, only a small subset of devices can be chosen for every global gradient aggregation because of the limitations in communication resources. To coordinate all edge devices, a scheduler is implemented at the edge server, which performs device selection and resource allocation.
\subsection{Learning Model}
In this work, we consider a general supervised machine learning task. We define an appropriate sample-wise loss function $\ell\left({\bf{w}}, {\bf{x}}, y\right)$ to quantify the prediction error between the data sample ${\bf{x}}$ on learning model $\bf{w}$ and the ground-truth label $y$. Specifically, Table I lists the loss functions for several popular learning models, which will be employed in the experiment.
\begin{table}
	\centering
	\small
	\caption{Loss Function for Popular Learning Models}
	\vspace{-1em}
	\label{table1}
	\begin{tabular}{p{4.5cm}|p{11cm}}
		\hline
		\hline
		Learning Model & Loss Function $\ell \left({\bf{w}}, {\bf{x}}, y\right)$ \\
		\cline{1-2}
		\hline
		\hline
		Least-squared Support Vector Machine (SVM) &$\frac{1}{2}\max\left\{0, 1-y {\bf{w}}^{\top}{\bf{x}}\right\} + \frac{\lambda}{2} \left \Vert {\bf{w}} \right \Vert ^2$, where $\lambda$ is a regularization parameter and $\top$ represents the transpose operator. \\
		\hline
		Linear Regression &$\frac{1}{2} \left\Vert y - {\bf{w}}^{\top}{\bf{x}} \right \Vert^2$\\
		\hline
        Neural Network &$\frac{1}{2}\left\Vert y - f\left({\bf{x}};{\bf{w}}\right) \right \Vert^2$, where $f\left({\bf{x}};{\bf{w}}\right)= f_N\left( \cdots f_2\left(f_1\left({\bf{x}};{\bf{w}}_1\right);{\bf{w}}_2\right)\cdots ;{\bf{w}}_N\right)$ is the learning output, $N$ is the number of layers, and $f_i \left({\bf{x}}, {\bf{w}}_i\right)$ is the $i$-th layer function conditioned on the weight matrix ${\bf{w}}_i$.\\
		%Neural Network &$\frac{1}{2} \left\Vert y - \sum_{i=1}^{I} \upsilon_i \phi\left({\bf{w}}^{\top}{\bf{x}}\right)\right \Vert^2$, where $I$ is the neuron number, $\upsilon_i$ is the weight connecting different neurons, and $\phi(\cdot)$ is the activation function.\\ %\cite{Deep_Learning}.\\
		\hline
		\hline
	\end{tabular}
\end{table}

Without loss of generality, we assume that the learning model is deployed across all edge devices. Then, the local loss function of device $k$ that measures the model error on its dataset $\mathcal{D}_k$ can be defined as
\begin{equation}\label{local loss function}
L_k\left({\bf{w}}\right)=\frac{1}{n_k} \sum_{i=1}^{n_k} \ell\left({\bf{w}},{\bf{x}}_k^i,y_k^i\right),~~~\forall k.
\end{equation}
Accordingly, the global loss function associated with all distributed local datasets is given by
\begin{equation}\label{global loss fucntion}
L\left({\bf{w}}\right)=  \frac{1}{n}\sum_{k=1}^K n_k L_k\left({\bf{w}}\right).
\end{equation}
The main objective of the learning task is to seek an optimal model ${\bf{w}}^*$ that minimizes $L\left({\bf{w}}\right)$. In addition, the ground-truth global gradient at ${\bf{w}}$ is defined as
\begin{equation}\label{ground-truth global gradient}
  {\bf{g}}=\nabla L\left({\bf{w}}\right),
\end{equation}
where $\nabla$ represents the gradient operator.

\subsection{Communication Model}
We assume orthogonal frequency-division multiple access (OFDMA) is adopted for uplink channel access. In this case, the system bandwidth, denoted by $B$, is divided into multiple narrowband sub-channels that are allocated to all devices without interference. Define $B_k$ as the bandwidth allocated to device $k$ so that we have $\sum_{k=1}^K B_k \leq B$. Moreover, we assume frequency flat fading, and let $\gamma_k$ denote the uplink signal-to-noise ratio (SNR) of device $k$, which is determined by its transmit power, uplink channel gain, and noise power. Then the achievable uplink data rate (in bit/s) for device $k$ can be expressed as
\begin{equation}\label{uplink data rate}
r_k= B_k \log_2 \left(1+ \gamma_k\right),~~~\forall k.
\end{equation}

For the downlink channel, we assume the edge server occupies the entire system bandwidth to broadcast the global model. Let $\gamma$ denote the minimum downlink channel SNR among the devices, which depends on their downlink channel gains, noise power, and the transmit power of the edge server. Then the achievable downlink data rate is given by
\begin{equation}\label{broadcast data rate}
r= B \log_2 \left(1+ \gamma \right).
\end{equation}

\subsection{Learning Mechanism}
\begin{figure}[htp]
	\centering
	\includegraphics[width=6.2in]{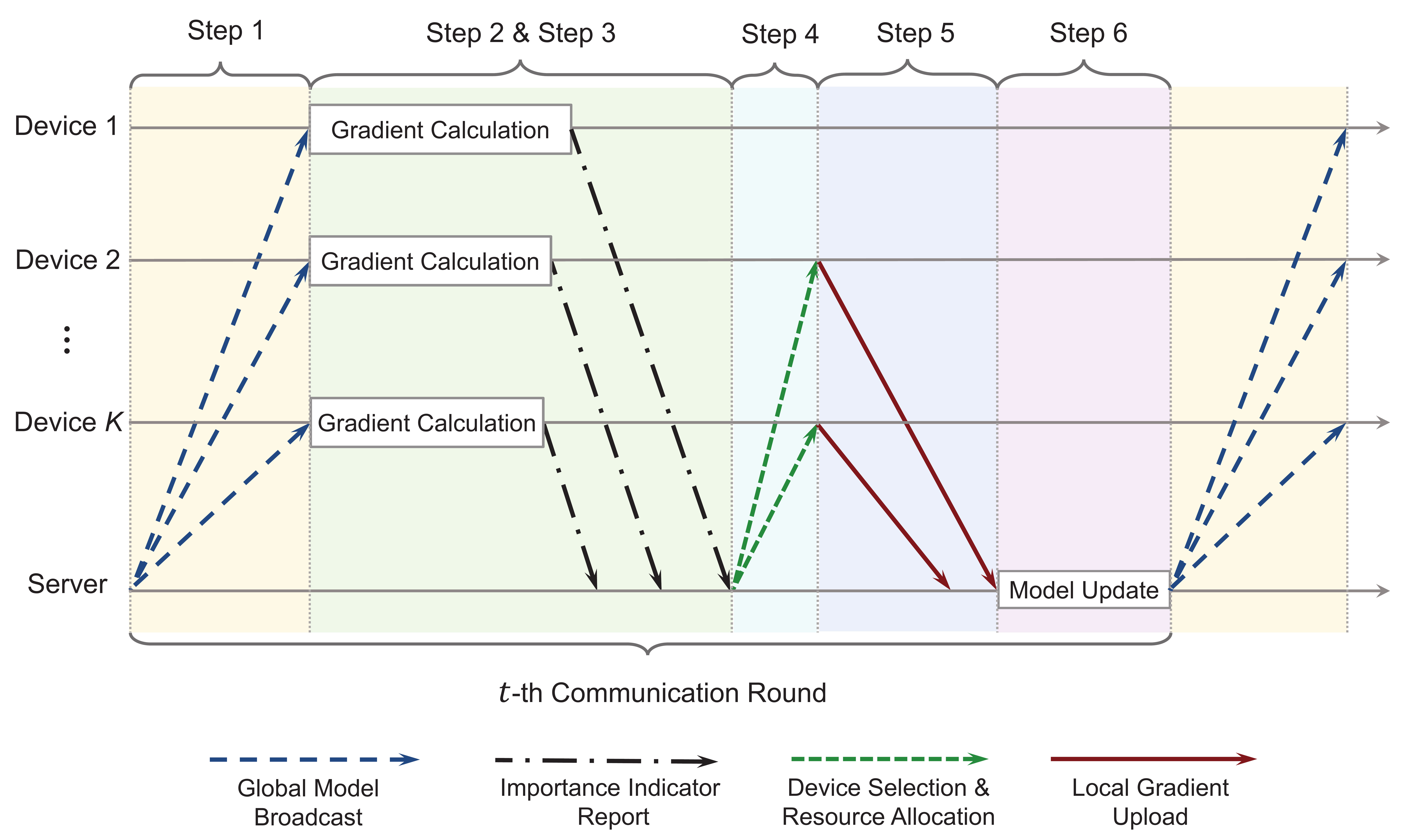}
	\caption{Communication protocol for importance- and channel-aware federated edge learning.}	
	\label{Protocol}
\end{figure}

We now introduce the federated learning mechanism by describing the steps that the communication parties take, as illustrated in Fig. \ref{Protocol}. The combination of the steps is referred to as a \textit{communication round}.
\begin{itemize}
    \item \textit{Step 1 (Global Model Broadcast):}
     The edge server broadcasts the current learning model ${\bf{w}}^t$ to all devices.
    %\item \textit{Step 2 (Local Gradient Calculation):}
%     Device $k$ updates its local gradient as
%     \begin{equation}\label{local gradient}
%          {\bf{g}}_k^t = \nabla L_k\left({\bf{w}}^t\right).
%        \end{equation}
    \item \textit{Step 2 (Local Gradient Calculation):}
     Device $k$ computes its local gradient as
     \begin{equation}\label{local gradient}
          {\bf{g}}_k^t = \nabla L_k\left({\bf{w}}^t\right).
        \end{equation}
	\item \textit{Step 3 (Importance Indicator Report):} As an option to be explained later, device $k$ computes an importance indicator $I\left({\bf{g}}_k^t\right)$ and reports the result to the edge server.
    \item \textit{Step 4 (Device Selection and Resource Allocation):} Based on the devices' reports, the edge server selects a subset of devices, denoted by $\mathcal{S}^t$, determines the bandwidth allocation, and then solicits updates from the selected devices via some control channels.
    \item \textit{Step 5 (Local Gradient Upload):} The selected devices upload their concrete local gradients to the edge server in the uplink multiaccess channel.\footnote{To facilitate theoretical analysis, we assume that the selected devices are always able to compute and upload their local gradients in the training duration.}
    \item \textit{Step 6 (Global Model Update):}
        After receiving all local gradients from the selected devices, the edge server computes the global gradient as
        \begin{equation}\label{global gradient}
          {{\bf{g}}}_{\mathcal{S}}^t =  \frac{\sum_{k \in \mathcal{S}^t}n_k {\bf{g}}_k^t}{\sum_{k \in \mathcal{S}^t} n_k}.
        \end{equation}
        Subsequently, the global model is updated to
        \begin{equation}\label{global update}
          {\bf{w}}^{t+1} = {\bf{w}}^{t} - \eta^t {{\bf{g}}}_{\mathcal{S}}^t,
        \end{equation}
            where $\eta^t > 0$ is the learning rate in the $t$-th communication round.
\end{itemize}
Starting from $t=1$, the edge server and edge devices iterate the preceding steps until convergence.

\section{Scheduling One Device in Each Round: System Design}
In this section, we consider a simple strategy of scheduling one device in each communication round. We first propose a new design principle and then develop a probabilistic scheduling framework. Thereafter, we devise a concrete gradient importance indicator. Based on these, we formulate the scheduling problem for the trade-off between multiuser channel and update diversity.
\subsection{Principle of Importance- and Channel-Aware Scheduling}
As discussed earlier, the main objective of the FEEL system is to minimize the training time for attaining a desired learning accuracy. According to the federated learning mechanism, we can observe that the training time is determined by two factors, i.e., the one-round latency and the model variation in each round. Both factors are affected by the scheduling decision, making device scheduling a key problem. Specifically, the former mainly depends on the channel fading due to that the communication latency dominates the computation latency \cite{Edge_Learning}. The latter is primarily decided by the selected local gradients based on the update equation (\ref{global update}). Since both factors are important, they should be balanced for joint scheduling design, leading to two types of multiuser diversity to be exploited in such a system. One is the independent fading in multiuser channels, namely \textit{multiuser channel diversity}, while the other is the heterogeneous importance levels across multiuser local gradients, namely \textit{multiuser update diversity}.
Targeting active gradient acquisition in FEEL, we should simultaneously exploit the multiuser diversity in wireless channels and local gradients, called \textit{channel-and-update diversity} to improve the learning performance.
\subsection{Probabilistic Scheduling}
In the classical federated averaging protocol, all devices are equally likely to participate in each communication round regardless of variability in the amount of data they possess. In such a way, the aggregated global gradient is generally biased \cite{FL_System_Design}. To address this issue, we now propose a new probabilistic scheduling framework.
%\textcolor[rgb]{0.00,0.00,1.00}{Let $\mathcal{P}^t = \left(p_1^t, p_2^t, \cdots, p_K^t\right)$ denote the \textit{scheduling distribution} at the edge server, where $p_k^t = P(X^t = k)$ is the probability that device $k$ can be selected in the $t$-th communication round.}
%For convenience, we first define a discrete random variable $X^t $ to denote the index of the scheduled device in the $t$-th communication round. Correspondingly, let $\mathcal{P}^t = \left(p_1^t, p_2^t, \cdots, p_K^t\right)$ denote the \textit{scheduling distribution} at the edge server, where $p_k^t$ is the probability that device $k$ can be selected in the $t$-th communication round.
%Then, the probabilistic scheduling framework can be defined as follows.
\begin{defi}[\textnormal{Probabilistic scheduling}]
In the $t$-th communication round, the edge server selects a random device $X^t \in \{1,2,\cdots,K\}$ for gradient uploading according to a scheduling distribution $\mathcal{P}^t = \left(p_1^t, p_2^t, \cdots, p_K^t\right)$ , where $p_k^t = P(X^t = k)$ is the probability that device $k$ can be selected. Upon receipt of the gradient update
${\bf{g}}_{X^t}^t$, the edge server computes the global gradient as
\begin{equation}\label{scaled global gradient}
\widehat{{\bf{g}}}^t = \dfrac{n_{X^t}}{n p_{X^t}^t}{\bf{g}}_{X^t}^t.
\end{equation}
\end{defi}

In this framework, probability $p_k^t$ can be interpreted as the level of importance that device $k$ can contribute to the global model convergence. %The design of $\mathcal{P}^t$ should jointly consider the multiuser channel and update diversity for optimal performance.
We shall note that before global model update, the selected local gradient ${\bf{g}}_{X^t}^t$ needs to be scaled by a coefficient $\dfrac{n_{X^t}}{n}$ at the edge server. This coefficient well quantifies the unbalanced property in global data distribution and thus makes the global gradient $\widehat{{\bf{g}}}^t$ unbiased.
\begin{lem}[\textnormal{Unbiasedness}]
The expectation of the global gradient defined in (\ref{scaled global gradient}) is equal to the ground-truth global gradient.
\end{lem}

\proof
Using (\ref{global loss fucntion}) and (\ref{ground-truth global gradient}), the ground-truth global gradient in the $t$-th communication round is given by
\begin{align}
{\bf{g}}^t &= \frac{\partial L\left( {\bf{w}}^t \right)}{ \partial{\bf{w}}^t} \notag\\
%&=\frac{1}{n} \frac{\partial \sum_{k=1}^K n_k L_k\left( {\bf{w}}^t \right)}{ \partial{\bf{w}}^t} \notag\\
&=\frac{1}{n} \sum_{k=1}^K n_k  \frac{\partial L_k\left({\bf{w}}^t \right)}{\partial{\bf{w}}^t} \notag\\
&=\frac{1}{n}\sum_{k=1}^K n_k {\bf{g}}_k^t. \label{ground-truth global gradient expression}
\end{align}
%\begin{align}
%{\bf{g}}^t &= \frac{\partial L\left( {\bf{w}}^t \right)}{ \partial{\bf{w}}^t} \notag\\
%&=\frac{1}{n} \frac{\partial \sum_{k=1}^K n_k L_k\left( {\bf{w}}^t \right)}{ \partial{\bf{w}}^t} \notag\\
%&=\frac{1}{n} \sum_{k=1}^K n_k  \frac{\partial L_k\left({\bf{w}}^t \right)}{\partial{\bf{w}}^t} \notag\\
%&=\frac{1}{n}\sum_{k=1}^K n_k {\bf{g}}_k^t. \label{ground-truth global gradient expression}
%\end{align}	
On the other hand, by taking expectation of $\widehat{\bf{g}}^t$ over the scheduling distribution $\mathcal{P}^t$, we have
\begin{align}
\mathbb{E}\left\{{\widehat{\bf{g}}^t}\right\} &= \sum_{k=1}^K p_k^t \dfrac{n_k}{n p_k^t}{\bf{g}}_k^t \notag\\
&= \frac{1}{n} \sum_{k=1}^K  n_k {\bf{g}}_k^t \notag\\
&= {\bf{g}}^t, \label{unbiasedness}
\end{align}
where (\ref{unbiasedness}) is due to (\ref{ground-truth global gradient expression}). Therefore, we can conclude that $\mathbb{E}\left\{{\widehat{\bf{g}}^t}\right\} = {\bf{g}}^t$, i.e., $\widehat{\bf{g}}^t$ is an unbiased estimate of the ground-truth global gradient.

The probabilistic scheduling yields unbiased gradient for all distributions as long as every device has a nonzero probability to update the server, including the typical uniform random scheduling in classical FL. However, different distributions lead to different performance in general.
\vspace{-1em}
\subsection{Gradient Importance Measurement}
To develop a concrete metric to evaluate the importance of each local gradient, we first analyze the contribution that each local gradient can provide for global model convergence. For the purpose of analysis, we make the following assumption to the global loss function.
\begin{assu}[\textnormal{Lipschitz gradient continuity}]
The gradient $\nabla L\left({\bf{w}}\right)$ of the global loss function $L\left({\bf{w}}\right)$ is Lipschitz continuous with a positive modulus $\ell$, or equivalently, for any $\bf{u}$ and $\bf{v}$, it satisfies
\begin{equation}\label{smooth assumption}
\left\Vert \nabla L\left({\bf{u}}\right) - \nabla L\left({\bf{v}}\right)\right\Vert \leq \ell \left\Vert {\bf{u}} - {\bf{v}}\right\Vert,
\end{equation}
where $\left\Vert \cdot \right\Vert$ represents the $L_2$ norm operator.
\end{assu}

Then, we give the convergence result of the probabilistic scheduling, as presented in the following lemma. The detailed proof is provided in Appendix A.
\begin{lem}[\textnormal{One-round convergence rate}]
When Assumptions 1 holds and let $\bf{w^*}$ denote the optimal global learning model. Then, the convergence rate of each communication round is given by
\begin{equation}\label{one round convergence rate}
\begin{aligned}
\mathbb{E}\left\{ L\left({\bf{w}}^{t+1}\right)-L\left(\bf{w^*}\right)\right\} \leq &~\mathbb{E}\left\{ L\left({\bf{w}}^{t}\right)-L\left(\bf{w^*}\right)\right\} - \eta^t \left(1 - \frac{1}{2} \ell \eta^t \right) \left\Vert {\bf{g}}^t \right\Vert^2  \\
&+ \frac{1}{2} \ell \left(\eta^t\right)^2 \mathbb{E}\left\{ \left \Vert\widehat{\bf{g}}^t - {\bf{g}}^t\right \Vert^2\right\}.
\end{aligned}
\end{equation}
\end{lem}

Lemma 2 reveals that the expected gap between the global loss value and the optimal one is bounded by the aggregate of three terms. The first term on the right-hand side of (\ref{one round convergence rate}) denotes the expected gap of the previous round. The second term is proportional to the squared norm of the ground-truth global gradient ${\bf{g}}^t$. The third term is proportional to the variance of the aggregated global gradient $\mathbb{E}\left\{ \left \Vert\widehat{\bf{g}}^t - {\bf{g}}^t\right \Vert^2\right\}$. The first two terms are independent of the scheduling decision such that they can be viewed as two constants. The variance of the aggregated global gradient depends on the scheduling design and thus needs to be optimized. Note that the ground-truth global gradient reflects the steepest direction for the global loss decline. Therefore, the smaller the variance of aggregated global gradient is, the faster the global loss decreases. %In addition, when our scheduling policy guarantees that the third term is smaller than the second term in each communication round, the expected gap between the global loss value and the optimal one will shrink as $t$ increases and the learning model will finally converge.

The result in Lemma 2 provides a significant connection between gradient variance and model convergence rate. Inspired by this, we can take the difference between the scaled local gradient and the ground-truth global gradient, called \textit{gradient divergence} as a reasonable importance measurement,  as described in the following definition.
\begin{defi}[\textnormal{Gradient divergence}]
\textnormal{Conditioned on the probabilistic scheduling framework, the importance indicator of each local gradient is defined as}
  \begin{equation}\label{Update Importance Indicator_S}
    I_k^t = \left \Vert \dfrac{n_k}{n p_k^t}{\bf{g}}_k^t - {\bf{g}}^t \right \Vert^2, ~~~\forall k.
  \end{equation}
\end{defi}

The gradient divergence reflects the deviation between each local gradient and the ground-truth global gradient. The smaller the local gradient divergence is, the more it can contribute to model convergence.
%The result in Lemma 2 provides a significant connection between gradient variance and model convergence rate. Inspired by this, we can take the difference between the scaled local gradient and the ground-truth global gradient, called \textit{gradient divergence} as a reasonable importance measurement,  as described in the following definition.
%\vspace{-0.7em}
%\begin{defi}[\textnormal{Gradient divergence}]
%Conditioned on the probabilistic scheduling framework, the importance indicator of each local gradient is defined as
%  \begin{equation}\label{Update Importance Indicator_S}
%    I_k^t = \left \Vert \dfrac{n_k}{n p_k^t}{\bf{g}}_k^t - {\bf{g}}^t \right \Vert^2, ~~~\forall k.
%  \end{equation}
%\end{defi}
%\vspace{-1em}
\subsection{One-Round Latency Analysis}
As mentioned earlier, the learning performance (training time) is determined not only by the gradient importance, but also by the one-round latency, making it an essential point to analyze. According to the previous federated learning mechanism, we can give the dominant kinds of latency in one communication round.
\begin{itemize}
  \item \textit{Global Model Broadcast Latency:} Let $S$ denote the total number of learning parameters and $q$ denote the quantitative bit number for each parameter. Then, the data size of the global model can be evaluated as $q \times S$ and the global model broadcast latency is given by
      \begin{align}\label{broadcasting latency}
            T^{\text{B},t} &= \dfrac{q S}{r^{t}} \notag\\
            &= \dfrac{q S}{B \log_2 \left(1+ \gamma^t \right) }.
      \end{align}
  \item \textit{Local Gradient Calculation Latency:} Define $C$ as the number of floating-point operations used for performing the backpropagation algorithm with one data sample. Moreover, let $f_k$ denote the computational capability (in floating-point operations per second) of device $k$. Accordingly, the local gradient calculation latency of device $k$ can be expressed as
      \begin{equation}\label{gradient calculation latency}
            T_k^{\text{C},t} = \dfrac{n_k C}{f_k},~~~\forall k.
      \end{equation}
  \item \textit{Local Gradient Upload Latency:} Due to the fact that each parameter has a counterpart gradient, the total number of elements in each local gradient equals $S$ as well. For convenience, we also use $q$ bits to quantize each gradient element. Besides, the selected device can occupy the entire system bandwidth since we schedule one device in each round. Given that device $k$ is scheduled, its local gradient upload latency can be expressed as
      \begin{align}\label{gradient uploading latency}
            T_k^{\text{U},t} &= \frac{q S}{r_k^{t}} \notag\\
            &=\frac{q S}{B \log_2 \left(1+\gamma_k^t\right)},~~~\forall k.
      \end{align}
\end{itemize}

Assuming that the edge server is computationally powerful, we ignore the time for device scheduling and global model update. Moreover, the edge server cannot start scheduling until it completes the channel state estimation and gradient importance evaluation for all devices. Towards this end, the one-round latency for scheduling device $k$ is given by
\begin{equation}
	T_k^t = T^{\text{B},t} + \max_{i \in \mathcal{K}} \left\{T_i^{\text{C},t} \right\} + T_k^{\text{U},t},~~~\forall k.
\end{equation}
\subsection{Problem Formulation}
Thus far, the multiuser channel and update diversity have been evaluated by the one-round latency and the gradient divergence, respectively. Targeting training acceleration, it is desirable to schedule the device with the best channel state as well as the smallest gradient divergence. However, this case rarely occurs in practice, resulting in a trade-off between gradient divergence and one-round latency. Define $\rho \in \left[0,1\right]$ as the weight coefficient that balances the gradient divergence and the one-round latency. \footnote{The selection of $\rho$ depends on the specific values of gradient divergence and one-round latency. A simple way is to choose a value of $\rho$ to make the weighted gradient divergence and the weighted one-round latency in the same order.} Then, the objective function can be defined as
\begin{align}
   \mathbb{E}\left\{\rho I_{X^t}^t  + (1-\rho) T_{X^t}^t  \right\} = \sum_{k=1}^K p_k^t \left( \rho \left \Vert \dfrac{n_k}{n p_k^t}{\bf{g}}_k^t - {\bf{g}}^t \right \Vert^2  + (1-\rho) T_k^t \right).\label{objective function}
\end{align}
Accordingly, the multiuser channel and update diversity trade-off problem can be formulated towards minimizing (\ref{objective function}), as
\begin{subequations}
	\begin{eqnarray}
	\mathscr{P}_1:~&\mathop{\text{minimize}}\limits_{\left(p_1^t,\cdots, p_K^t\right)} &\sum_{k=1}^K p_k^t \left( \rho \left \Vert \dfrac{n_k}{n p_k^t}{\bf{g}}_k^t - {\bf{g}}^t \right \Vert^2  + (1-\rho) T_k^t \right),\label{P1a}\\
	&{\text{subject to}}&\sum_{k=1}^K p_k^t = 1,\label{P1b}\\
    && p_k^t \geq 0,~ \forall k \in \mathcal{K}. \label{P1c}
	\end{eqnarray}
\end{subequations}

\section{Scheduling Optimization and Convergence Analysis}
In this section, we first develop the optimal scheduling policy to Problem $\mathscr{P}_1$. Based on this, we will provide the convergence performance of the FEEL system.
\subsection{Optimal Scheduling Policy}
The main challenge in solving Problem $\mathscr{P}_1$ is that the ground-truth global gradient cannot be obtained by neither the edge server nor edge devices, making the gradient divergence hard to calculate in practice. To make progress, we first rewrite the first term in the objective function (\ref{objective function}) as
\begin{align}
\mathbb{E}\left\{\rho I_{X^t}^t \right\} &=\sum_{k=1}^K p_k^t \left( \rho \left \Vert \dfrac{n_k}{n p_k^t}{\bf{g}}_k^t - {\bf{g}}^t \right \Vert^2 \right) \notag\\
&= \sum_{k=1}^K \left[\frac{\rho}{p_k^t} \left(\frac{n_k}{n}\right)^2 \left\Vert{\bf{g}}_k^t\right\Vert^2 \right] - \rho \left\Vert{\bf{g}}^t\right\Vert^2, \label{progress}
\end{align}
where (\ref{progress}) is due to the unbiasedness property of the aggregated global gradient. In particular, the ground-truth global gradient, the global model broadcast latency, and the local gradient calculation latency are independent of the scheduling decision so that they can be excluded from the optimization problem. Accordingly, Problem $\mathscr{P}_1$ can be equivalently transformed into
\begin{align} \mathscr{P}_2:~&\mathop{\text{minimize}}\limits_{\left(p_1^t,\cdots, p_K^t\right)} ~~\sum_{k=1}^K \left[\frac{\rho}{p_k^t} \left(\frac{n_k}{n}\right)^2 \left\Vert{\bf{g}}_k^t\right\Vert^2 + (1-\rho) p_k^t T_k^{\text{U},t} \right],\label{P2a}\\
	&{\text{subject to}} ~~~\text{(\ref{P1b}) and (\ref{P1c})}. \notag%\label{P2b}
\end{align}

To solve Problem $\mathscr{P}_2$, we first establish the following lemma, which is proved in Appendix B.
%\vspace{-0.6em}
\begin{lem}
Given a general optimization problem with positive coefficients $a_k$, $b_k$, and $c_k$ as follows
\begin{subequations}
\begin{eqnarray} \mathscr{P}_3:&\mathop{\text{minimize}}\limits_{\left(x_1,\cdots, x_K\right)} &\sum_{k=1}^K a_k \frac{1}{x_k} +b_k x_k,\label{P3a}\\
	&{\text{subject to}} &\sum_{k=1}^{K} c_k x_k = d,\label{P3b}\\
&&x_k \geq 0,~ \forall k \in \mathcal{K}. \label{P3c}
	\end{eqnarray}
\end{subequations}
The optimal solution can be expressed as
\begin{equation}\label{solution}
	x_k^*= \sqrt{\frac{a_k}{b_k+\lambda^*c_k}},~~~\forall k.
\end{equation}
where $\lambda^*$ is the optimal value satisfying $\sum_{k=1}^{K} c_k x_k^* = d$.
\end{lem}

%\vspace{-0.5em}
With comprehensive comparison, we can observe that the structures of problems $\mathscr{P}_2$ and $\mathscr{P}_3$ are identical such that the result in (\ref{solution}) is applicable in solving $\mathscr{P}_2$. Towards this end, we can obtain the optimal scheduling policy, as characterized in the following theorem.
%\vspace{-0.65em}
\begin{thm}[\textnormal{Importance- and channel-aware scheduling}]
Considering the strategy of scheduling one device in each communication round, the optimal probability that the edge server selects device $k$ to report its local gradient is given by
\begin{equation}\label{scheduling probability of one device}
	p_k^{t*}= \frac{n_k}{n} \left\Vert {\bf{g}}^t_k \right\Vert \sqrt{ \frac{\rho}{(1-\rho)T_k^{\text{U},t} + \lambda^{t*}}},~~~\forall k,
\end{equation}
where $\lambda^{t*}$ is the optimal value of the Lagrangian multiplier that guarantees $\sum_{k=1}^K p_k^{t*} = 1$ and can be obtained by the one-dimensional searching algorithm.
\end{thm}

%\vspace{-0.4em}
Theorem 1 reveals that the optimal scheduling decision is mainly determined by three factors: the data unbalanced indicator $\frac{n_k}{n}$, the local gradient norm $\left\Vert{\bf{g}}_k^t\right\Vert$, and the local gradient upload latency $T_k^{\text{U},t}$. More precisely, for given value of $\lambda^{t*}$, the scheduling probability increases linearly with both the data unbalanced indicator and the local gradient norm, whereas decreases sublinearly with the exponent of $-\frac{1}{2}$ when the local gradient upload latency is large.
%Theorem 1 reveals that the optimal scheduling decision is mainly determined by three factors: the data unbalanced indicator $\frac{n_k}{n}$, the local gradient norm $\left\Vert{\bf{g}}_k^t\right\Vert$, and the local gradient upload latency $T_k^{\text{U},t}$. More precisely, the scheduling probability increases linearly with both the data unbalanced indicator and the local gradient norm, whereas decreases with the local gradient upload latency approximately in the order of $-\frac{1}{2}$.
The data unbalanced indicator reflects the data size proportion of each local dataset in the global dataset. Therefore, a larger data unbalanced indicator implies that the local gradient contains more data information such that it can contribute more to the model convergence. On the other hand, it is desirable to schedule the device with a larger gradient norm as it includes more non-zero elements and can further promote the model update.
Moreover, the result suggests us to assign a small probability to the device with bad channel state. Also, scheduling a device with bad channel state will incur straggler issue, which should be avoided in the scheduling design.
%Last but not least, the result suggests us to assign a small probability to the device with bad channel state, which is consistent with the intuition.

From Theorem 1, we can observe that the importance level of each local gradient is well characterized by two scalars: its gradient norm and local dataset size. Moreover, both scalars can be easily derived at local devices while transmitting them to the scheduler will not incur excessive delay. This result drives us to use the product of these two scalars as the gradient importance measurement. On the other hand, the scheduler is able to evaluate each local gradient upload latency via dynamical channel state estimation. Towards this end, both channel and update diversity can be jointly exploited for practical scheduling design.

\subsection{Convergence Analysis}
%We now investigate the convergence behaviour of the learning algorithm under the proposed scheduling solution. By applying the result of Theorem 1 into Lemma 2, we can derive the convergence upper bound of each communication round, as shown in the following theorem. The detailed proof is presented in Appendix C.
We now investigate the convergence behaviour of the learning algorithm under the proposed scheduling solution. To facilitate analysis, we further make the following assumption to the global loss function.
\begin{assu}[\textnormal{Strong convexity}]
The global loss function $L\left({\bf{w}}\right)$ is strongly convex with a positive parameter $\mu$ such that for any $\bf{u}$ and $\bf{v}$, it follows
\begin{equation}
L\left(\bf{u}\right) \geq  L\left(\bf{v}\right) + \nabla L\left(\bf{v}\right)^\top \left( \bf{u} - \bf{v}\right) + \frac{\mu}{2} \left\Vert \bf{u} - \bf{v} \right\Vert^2.  \label{convex assumption}
\end{equation}
\end{assu}

This standard assumption is satisfied by many popular learning models, such as the least-squared SVM and the linear regression in Table I. In Section VI, we will conduct experiments to demonstrate that our analytical results also work well for some popular loss functions which do not satisfy this assumption. Then, by applying the result of Theorem 1 into Lemma 2, we can derive the convergence upper bound of each communication round, as shown in the following theorem. The detailed proof is presented in Appendix C.
%\vspace{-0.7em}
\begin{thm}[\textnormal{Convergence upper bound}]
	Conditioned on the scheduling policy in Theorem 1, the convergence upper bound in the $t$-th communication round can be given by
\begin{equation}
\begin{aligned}\label{Convergence upper bound}
	\mathbb{E}\left\{ L\left({\bf{w}}^{t+1}\right) - L\left(\bf{w^*}\right)\right\}& \leq \prod_{i=1}^t \left(1-2\mu \eta^i\right) \mathbb{E}\left\{ L\left({\bf{w}}^1\right)-L\left(\bf{w^*}\right)\right\}\\
	& + \frac{\ell}{2 n} \sum_{i=1}^t A^i \left(\eta^i\right)^2 \sum_{k=1}^{K} n_k \left\Vert {\bf{g}}_k^i\right\Vert \sqrt{\dfrac{\left(1-\rho\right)T_k^{\text{U},i}+\lambda^{i*}}{\rho}},
    \end{aligned}
\end{equation}
where $A^i=\prod_{j=i+1}^{t} \left(1-2\mu \eta^j\right)$ is a weight coefficient.
\end{thm}

Theorem 2 indicates that the expected gap between the global loss value and the optimal one is bounded by the aggregate of two terms: one is the expected gap in the initial round and the other is the cumulate impact of the proposed scheduling policy on model convergence. The former will converge to zero as $t$ increases when the learning rate satisfies $0< \eta^i <\frac{1}{2\mu}$, $\forall i$. Meanwhile, the weight coefficient $A^i$ will also decrease, making the latter approach to zero as well. Therefore, the learning model will finally converge to the optimum under the proposed scheduling policy.

The convergence upper bound in Theorem 2 holds for any learning rate that satisfies $0< \eta^i <\frac{1}{2\mu}$ $\forall i$. To explicitly characterize the convergence performance, we diminish the learning rate as the training proceeds such that the convergence rate of the learning algorithm can be given as follows. The detailed proof is provided in Appendix D.
\vspace{-0.55em}
\begin{coro}[\textnormal{Learning convergence rate}]
    Let $T$ denote the number of communication rounds, and $\eta^t = \dfrac{\chi}{t+\nu}$ denote the learning rate for given $\chi > \dfrac{1}{2 \mu}$ and $\nu > 0$. Then the convergence rate of the learning algorithm is given by
    \begin{equation}\label{Specific Convergence Rate}
      \mathbb{E}\left\{ L\left({\bf{w}}^T\right)-L\left(\bf{w^*}\right)\right\} \leq \dfrac{\zeta}{T+\nu},
    \end{equation}
    where $\zeta = \max \left\{ \dfrac{\ell G^2 \chi^2}{2\left(2 \mu \chi -1 \right)},\left(1 + \nu\right)\mathbb{E}\left\{ L\left({\bf{w}}^1\right)-L\left(\bf{w^*}\right)\right\}\right\}$ and $G = \max \left\Vert \widehat{\bf{g}}^t \right\Vert $ is the largest gradient norm across the communication rounds.
\end{coro}

%\vspace{-0.7em}
According to Corollary 1, we can conclude that the increment of total communication rounds leads to the convergence of the learning algorithm. Moreover, the convergence rate is asymptotically $\mathcal{O}\left(\frac{1}{T}\right)$.
\section{Scheduling Multiple Devices in Each Round}
In this section, we investigate the strategy of scheduling multiple devices in each communication round. We first develop a low-complexity algorithm for scheduling design and then propose an optimal bandwidth allocation for practical implementation.
%\vspace{-0.3em}
\subsection{Scheduling Design}
The probabilistic scheduling framework is still applicable in the strategy of scheduling multiple devices in each communication round by combining the \textit{sampling without replacement} approach \cite{PKU}. Let $M$ denote the number of devices that need to be scheduled in each round. Then the probabilistic scheduling framework can be defined as follows.
\begin{defi}[\textnormal{Probabilistic scheduling without replacement}]
In the $t$-th communication round, the edge server selects a random device sequence $\mathcal{Y}^t$ = $\left(Y_1^t, Y_2^t, \cdots, Y_M^t\right)$ without replacement according to the scheduling distribution $\mathcal{P}^t = \left(p_1^t, p_2^t, \cdots, p_K^t\right)$, where $Y_m^t \in \{1,2,\cdots, K\}$ is a random variable representing the index of the $m$-th scheduled device. Upon receipt of the gradient updates ${\bf{g}}_{Y_m^t}^t$, the edge server computes the global gradient as
\begin{equation}\label{global gradient in multiple case}
  \widehat{{\bf{g}}}_{\mathcal{Y}}^t = \frac{1}{M n} \sum_{m = 1}^M \dfrac{n_{Y_m^t}}{q_{Y_m^t}^t} {\bf{g}}_{Y_m^t}^t,
\end{equation}
where $q_{Y_m^t}^t = \frac{p_{Y_m^t}^t}{1-\sum_{j=1}^{m-1} p_{Y_j^t}^t}$ is the probability of scheduling device $Y_m^t$ conditioned on the selected device sequence $\left(Y_1^t,\cdots, Y_{m-1}^t \right)$.
\end{defi}
%\vspace{-1.5em}
\begin{lem}[\textnormal{Unbiasedness}]
The expectation of the global gradient defined in (\ref{global gradient in multiple case}) is equal to the ground-truth global gradient.
\end{lem}

%\vspace{-1em}
\proof
The proof can be conducted by proving the unbiasedness of each selected local gradient. Assume that we have selected $m-1$ devices, denoted by a sequence $\left(k_1^t, \cdots, k_{m-1}^t \right)$. Then the scheduling probabilities for these devices become zero in the $m$-th selection step. Accordingly, the conditional scheduling distribution for selecting the next device is given by
\begin{equation}\label{new probability}
q_{k}^t = \left \{
    \begin{aligned}
        &0,~~&&\text{if}~~ k \in \left(k_1^t,\cdots,k_{m-1}^t \right),\\
        &\frac{p_{k}^t}{1-\sum_{j=1}^{m-1} p_{k_j^t}^t},~~&&\text{otherwise}.
	\end{aligned} \right.
\end{equation}
Consequently, the edge server randomly selects a device $Y_m^t$ and computes the scaled local gradient by
\begin{equation}\label{scaled local gradient in multiple devices}
    \widehat{{\bf{g}}}_{Y_m^t}^t= \dfrac{n_{Y_m^t}}{n q_{Y_m^t}^t}{\bf{g}}_{Y_m^t}^t.
\end{equation}
Taking expectation of $\widehat{{\bf{g}}}_{Y_m^t}^t$ over the conditional scheduling distribution, we have
\begin{align}
\mathbb{E}\left\{\widehat{{\bf{g}}}_{Y_m^t}^t\right\} &= \sum_{k=1}^K q_k^t \dfrac{n_k}{n q_k^t}{\bf{g}}_k^t \notag\\
&= \frac{1}{n} \sum_{k=1}^K  n_k {\bf{g}}_k^t \notag\\
&= {\bf{g}}^t. \label{unbiasedness of multiple gradients}
\end{align}
Therefore, each selected local gradient is unbiased, making their average defined in (\ref{global gradient in multiple case}) an unbiased estimate of the ground-truth global gradient. This completes the proof.
%\end{IEEEproof}

Based on the preceding discussion, we now present an efficient scheduling algorithm, as described in Algorithm 1. To facilitate practical implementation, we first assume that the entire bandwidth is available to each device and then sequentially select one device from the candidate device set in each iteration until reaching the required number of devices. Although it sacrifices the optimality but will significantly reduce the computational complexity. Specifically, the total number of iterations is $M$. Moreover, let $\epsilon$ denote the maximum tolerance for the one-dimensional search for $\lambda^{t*}$. Then, the computational complexity of Algorithm 1 is $\mathcal{O}\left( \log\frac{1}{\epsilon} + M\right)$, which can be easily implemented in practical systems.

%Based on the above discussion, we now present an efficient scheduling algorithm, as described in Algorithm 1. The main idea of the algorithm is to sequentially select one device from the candidate device set in each iteration until reaching the required number of devices. Moreover, let $\epsilon$ denote the maximum tolerance for the one-dimensional search for $\lambda^{t*}$. Then, the computational complexity of Algorithm 1 is $\mathcal{O}\left( \log\frac{1}{\epsilon} + M\right)$, which can be easily implemented in practical systems.

\begin{algorithm}
\caption{Importance- and Channel-Aware Scheduling at Edge Server}\label{algorithm}
\KwIn{Uplink SNR $\gamma_k^t$, local gradient norm $\left\Vert {\bf{g}}^t_k \right\Vert$, local dataset size $n_k$, scheduled device number $M$.}
\KwOut{Scheduling device sequence $\mathcal{Y}^t$.}
Initialize $\mathcal{Y}^t = \emptyset$, $\mathcal{Q}^t = \emptyset$, $m=1$.\\
Compute the optimal scheduling distribution $\mathcal{P}^{t*} = \left(p_1^{t*}, p_2^{t*}, \cdots, p_K^{t*}\right)$ according to (\ref{scheduling probability of one device}), involving the one-dimensional search for $\lambda^{t*}$.\\
Let $\mathcal{Q}^t \leftarrow \mathcal{P}^{t*}$.\\
\Repeat{$m>M$}
{Randomly select a device $k_m^t$ based on $\mathcal{Q}^{t}$.\\
$\mathcal{Y}^t \leftarrow \mathcal{Y}^t + \left\{k_m^t\right\}$.\\
Update $\mathcal{Q}^{t}$ according to (\ref{new probability}).\\
$m \leftarrow  m+1$.}
\end{algorithm}
\vspace{-2em}
\subsection{Bandwidth Allocation}
Given a typical device sequence $\mathcal{Y}^t = \left(k_1^t,k_2^t,\cdots,k_M^t\right)$ for gradient aggregation, another concern related is the corresponding bandwidth allocation towards minimizing the one-round latency. Let $B_{k_m^t}^t$ denote the bandwidth allocated to the device $k_m^t$ in the $t$-th communication round. Then, the local gradient upload latency can be expressed as
\begin{align}\label{gradient uploading latency multiple devices}
T_{k_m^t}^{\text{U},t} &= \dfrac{q S}{r_{k_m^t}^{t}} \notag\\
&= \frac{q S}{B_{k_m^t}^t \log_2 \left(1+ \gamma_{k_m^t}^t\right)},~~~\forall k_m^t \in \mathcal{Y}_t.
\end{align}
On the other hand, the global model broadcast latency and the local gradient calculation latency are both independent of the bandwidth allocation. Thus, the bandwidth allocation should be optimized towards minimizing the overall gradient upload latency, as
\begin{subequations}
	\begin{eqnarray} \mathscr{P}_4:&\mathop{\text{minimize}}\limits_{\left(B_{k_1^t}^t, \cdots, B_{k_M^t}^t \right)} & \mathop{\text{maximize}}\limits_{k_m^t \in \mathcal{Y}^t} \left\{\frac{q S}{B_{k_m^t}^t \log_2 \left(1+ \gamma_{k_m^t}^t\right)}\right\}, \label{P4a}\\
	&{\text{subject to}}&\sum_{k_m^t \in \mathcal{Y}^t} B_{k_m^t}^t \leq B,\label{P4b}\\
    &&B_{k_m^t}^t \geq 0,~\forall k_m^t \in \mathcal{Y}^t.\label{P4c}
	\end{eqnarray}
\end{subequations}

It is easy to observe that $\mathscr{P}_4$ is convex such that it can be solved by the KKT conditions \cite{KKT}. For ease of notation, define $R_{k_m^t}^t = \log_2 \left(1+\gamma_{k_m^t}^t\right)$ as the instantaneous uplink data rate of device $k_m^t$ with unit bandwidth. Let $\left(B_{k_1^t}^{t*},B_{k_2^t}^{t*}, \cdots, B_{k_M^t}^{t*} \right)$ denote the optimal solution to $\mathscr{P}_4$. Then, by parametric algorithm and simple mathematical calculation, we can find that the optimal bandwidth allocation is achieved when the local gradient upload latency for each device is equalized. Therefore, the optimal bandwidth allocation can be presented in the following theorem.
%It is easy to observe that $\mathscr{P}_4$ is convex such that it can be solved by the KKT conditions. For ease of notation, define $R_{k_m^t}^t = \log_2 \left(1+\gamma_{k_m^t}^t\right)$ as the instantaneous uplink data rate of device $k_m^t$ with unit bandwidth. Let $\left(B_{k_1^t}^{t*},B_{k_2^t}^{t*}, \cdots, B_{k_M^t}^{t*} \right)$ denote the optimal solution to $\mathscr{P}_4$. Then, we can present the optimal bandwidth allocation in the following theorem. The detailed proof is omitted due to page limits.
%\vspace{-0.75em}
\begin{thm}[\textnormal{Bandwidth allocation}] \!
Given the device sequence $\mathcal{Y}^t = \left(k_1^t, k_2^t, \cdots, k_M^t\right)$ to report their local gradients in the $t$-th communication round, the optimal bandwidth allocation is given by
  \begin{equation}\label{bandwdith allocation}
    B_{k_m^t}^{t*} = \frac{B}{R_{k_m^t}^t}\left(\sum_{m=1}^M \frac{1}{R_{k_m^t}^{t}}\right)^{-1},~~~\forall k_m^t \in \mathcal{Y}_t.
  \end{equation}
\end{thm}

The result in Theorem 3 is rather intuitive that more bandwidth should be allocated to the devices with worse channel states to achieve the smallest one-round latency. By this means, the communication latency of each scheduled device can be equalized and thus facilitate the update synchronization needed for gradient aggregation.
\section{Numerical Results}
In this section, we conduct experiments to validate the theoretical analysis and test the performance of the proposed algorithms. All codes are implemented in python 3.6 and are run on a Linux server equipped with four NVIDIA GeForce GTX 1080 Ti GPUs.
%\vspace{-0.55em}
\subsection{Experiment Settings}
%\vspace{-0.3em}
The default experiment settings are given as follows unless specified otherwise. We consider a small-cell network with one edge server and $K=30$ edge devices. The cell radius is $500$ m. The edge server is located at the center of the network and each device is uniformly distributed within the coverage. The path loss between each device and the edge server is generated by $128.1 + 37.6\log_{10}(d)$ (in dB) according to the LTE standard \cite{LTE}, where $d$ is the device-to-server distance in kilometer. The channel noise power density is $-174$ dBm/Hz. The transmit powers of each device and the edge server are set as $24$ dBm and $46$ dBm, respectively. The system bandwidth is $B= 1$ MHz. The average quantitative bit number for each gradient and parameter element is $q=16$ bits.

For exposition, we consider the learning task of training classifiers. Two prevalent learning models of least-squared SVM and convolutional neural network (CNN) are employed for implementation.\footnote{The CNN has 6 layers including two $5\times5$ convolution layers with ReLu activation (the first layer with 32 channels and the second layer with 64 channels), each followed with a $2\times2$ max pooling layer, a fully connected layer with 512 units and ReLu activation, and a softmax output layer.} To better simulate the mobile data distribution in FEEL, we consider the non-IID data partitioning way as follows. For the SVM model, we choose two typical classes of ``airplane" and ``automobile" in the well-known dataset CIFAR-10 for classification, where each device owns 330 data samples in one class. For the CNN model, we use the popular MNIST dataset that consists of 10 categories ranging from digit ``0" to ``9" and a total of 60,000 labeled training data samples. We first sort all data samples by their digit labels, divide them into 60 shards of size 1,000, and then assign each device with two shards. By this means, each device obtains the data samples with only two types of digits, making the data distribution over devices a pathological non-IID manner. Moreover, the learning step-sizes for the SVM model and CNN model are set as $\eta_{\text{SVM}} = 0.0001$ and $\eta_{\text{CNN}} = 0.005$, respectively.

To demonstrate the effectiveness of the proposed scheduling policy, two baseline policies namely \textit{channel-aware scheduling} and \textit{importance-aware scheduling} are also implemented in the following experiments. Specifically, the coefficient is $\rho=0$ in the former such that the scheduling decision depends merely on the channel state. Meanwhile, the coefficient is $\rho=1$ in the latter such that the scheduling decision is solely decided by the update importance.
%To demonstrate the effectiveness of the proposed scheduling policy, two baseline policies namely \textit{channel-aware scheduling} and \textit{importance-aware scheduling} are also implemented in the following experiments. The scheduling decision in the former depends merely on the channel state while the scheduling decision in the latter is solely decided by the update importance. We name our proposed policy as \textit{importance- and channel-aware scheduling} in the sequel.
\subsection{Scheduling One Device in Each Round}
\begin{figure}
	\centering
	\subfigure[SVM]{
		\includegraphics[width=3.7in]{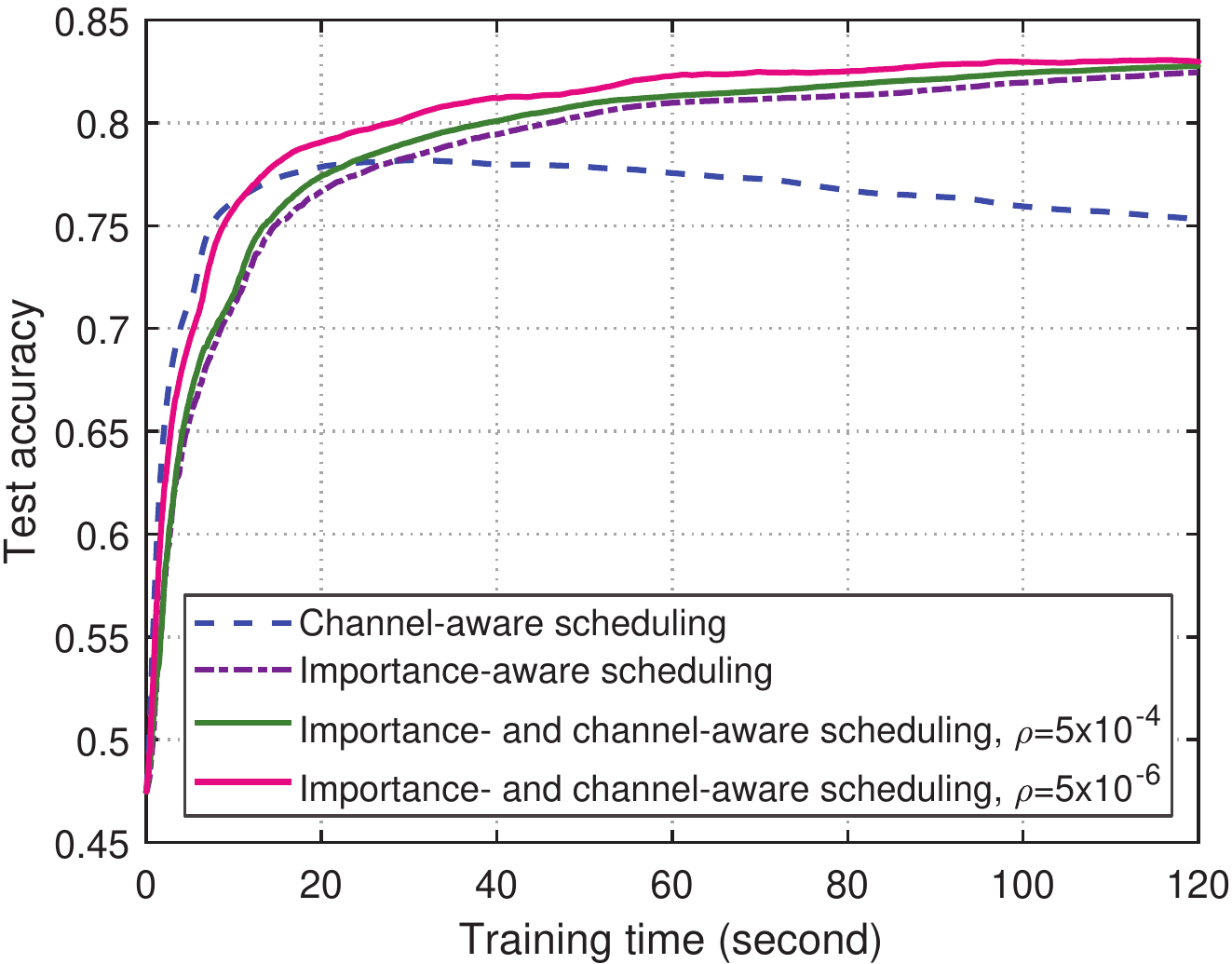}
        \label{SVM_Scheduling one user each round}
	}
	\subfigure[CNN]{
		\includegraphics[width=3.7in]{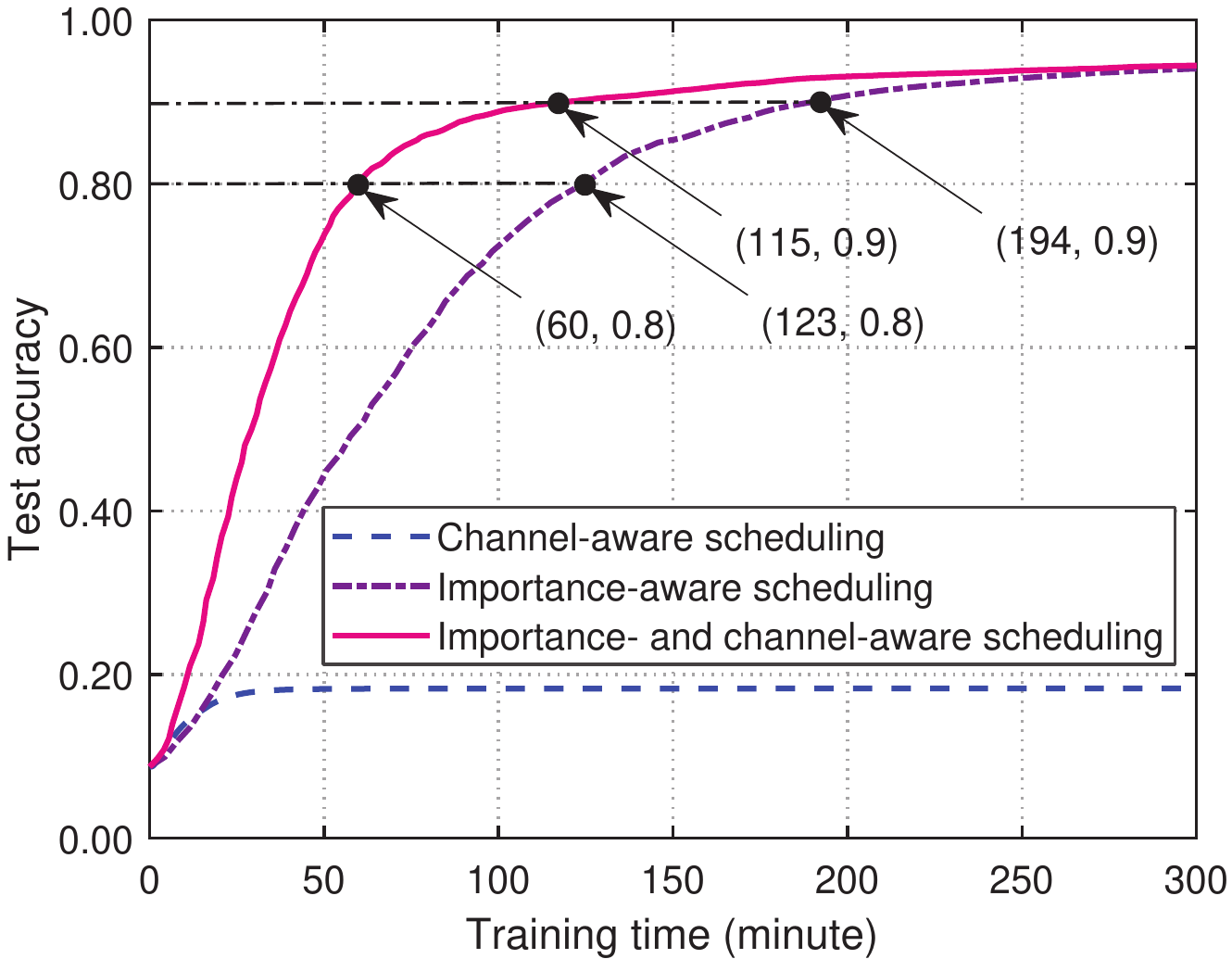}
        \label{CNN_Scheduling one user each round}
	}
	\caption{Performance comparison among different scheduling policies in the strategy of scheduling one device in each communication round.}
\end{figure}
\textit{1) Performance Comparison in SVM:} Fig. \ref{SVM_Scheduling one user each round} depicts the learning performance of the proposed scheduling policy as compared with the two baseline policies using the SVM model. From this figure, we can observe that at the initial stage, the proposed policy with $\rho=5\times 10^{-6}$ can achieve comparable convergence speed as that of the channel-aware scheduling but converges faster than the importance-aware scheduling. It is because that the learning updates of different devices play a similar role in model convergence at the beginning of the training process, making channel state the dominant factor for learning performance improvement. Nevertheless, the proposed policy can outperform the channel-aware scheduling as the training proceeds. The underlaying reason is that the gain by exploiting channel diversity is saturated such that the improvement primarily comes from the update diversity. In particular, the channel-aware scheduling suffers from a visible performance degradation after achieving an accuracy peak. As we can imagine, part of devices with poor communication channels cannot participate in the training process when scheduling decision merely depends on the channel state. This inevitably incurs the data deficiency and will result in overfitting issue. On the other hand, the proposed policy always outperforms the importance-aware scheduling policy since it well balances the channel state and update importance in the training duration. Last, we can find that the weight coefficient $\rho$ has a great impact on model convergence. In specific, an overlarge $\rho$ may lead to a marginal performance gain in that the proposed policy is hardly to exploit the channel diversity across edge devices. This suggests that the coefficient should be carefully picked to balance the channel state and update importance and thereby improving the learning performance.

\textit{2) Performance Comparison in CNN:} Fig. \ref{CNN_Scheduling one user each round} illustrates the performance comparison among the three scheduling policies using the CNN model. The weight coefficient for the proposed scheduling policy is set as $\rho=5 \times 10^{-3}$. Similar trends as in the SVM model are observed, and the proposed policy is found to consistently surpass the baseline policies, which confirms the robustness and stability of the theoretical results against the model structure. More precisely, when the target accuracy is $0.8$, the required training time is $60$ minutes for the importance- and channel-aware scheduling while the importance-aware scheduling takes 123 minutes to achieve the same accuracy. Therefore, the proposed policy can save more than half time to achieve the target accuracy comparing against the importance-aware scheduling. The performance gain becomes more remarkable if the target accuracy is $0.9$. In contrast, the channel-aware scheduling is incapable of achieving the target accuracy due to the lack of data information from the devices with poor channel conditions. In particular, the performance gap between the channel-aware scheduling and importance- and channel-aware scheduling is more evident than that in the SVM model. It attributes to the fact that the update diversity in the multiclass classification scenario is larger than that in the binary classification scenario. These observations further demonstrate the superiority of the proposed scheduling policy to deal with the non-IID data in distributed FEEL system.
\begin{figure}[htp!]
	\centering
	\subfigure[SVM]{
		\includegraphics[width=3.7in]{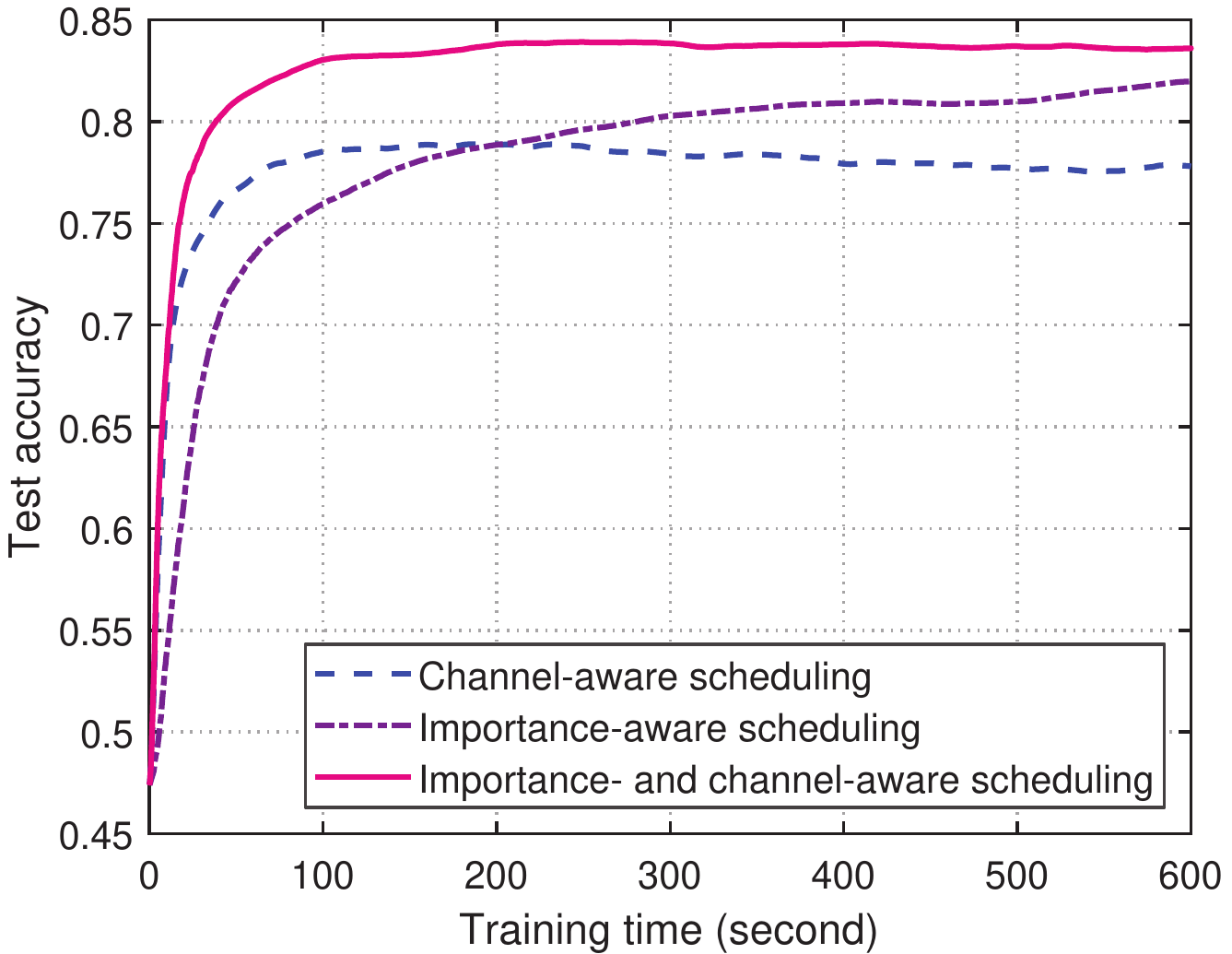}
        \label{SVM_Scheduling 10 user each round}
	}
	\subfigure[CNN]{
		\includegraphics[width=3.7in]{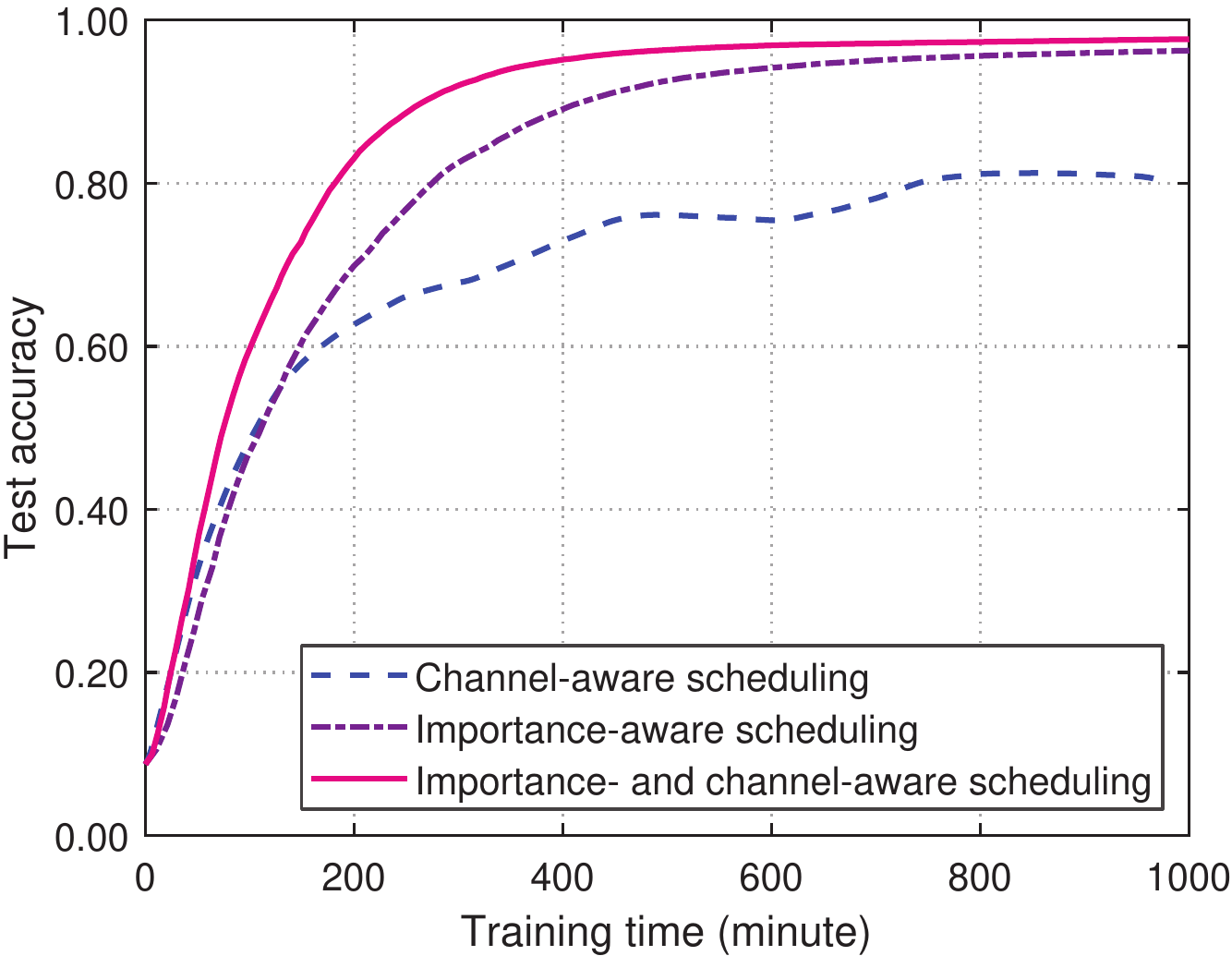}
         \label{CNN_Scheduling 3 user each round}
	}
	\caption{Performance comparison among different scheduling policies in the strategy of scheduling multiple devices in each communication round.}
\end{figure}
\subsection{Scheduling Multiple Devices in Each Round}
\textit{1) Performance Comparison:} The learning performance of the three scheduling policies are evaluated in the strategy of scheduling multiple devices in each communication round. Both the models of SVM and CNN are experimented. For SVM, we assume that $M=10$ devices can be scheduled in each round and the coefficient is set as $\rho= 5 \times 10^{-6}$. For CNN, we assume that $M=3$ devices can be scheduled in each round and the coefficient is set as $\rho= 1 \times 10^{-3}$. The results are displayed in Figs. \ref{SVM_Scheduling 10 user each round} and \ref{CNN_Scheduling 3 user each round}, respectively. It can be observed from all plots that the proposed policy achieves the fastest convergence speed along with the highest learning accuracy throughout the entire training duration. This result is rather intuitive since the channel-aware scheduling overlooks the update importance whereas the importance-aware scheduling is unconscious of the channel condition. It also confirms the applicability of the proposed sequential scheduling without replacement algorithm, even though it cannot guarantee the optimal device scheduling. On the other hand, the performance improvement in this strategy is more conspicuous than that in the strategy of scheduling one device in each communication round, which substantiates the theoretical gain brought by the intelligent trade-off between quality and quantity in update aggregation.

\begin{figure}[htp!]
	\centering
	\subfigure[$B=1$ MHz]{
		\includegraphics[width=3.7in]{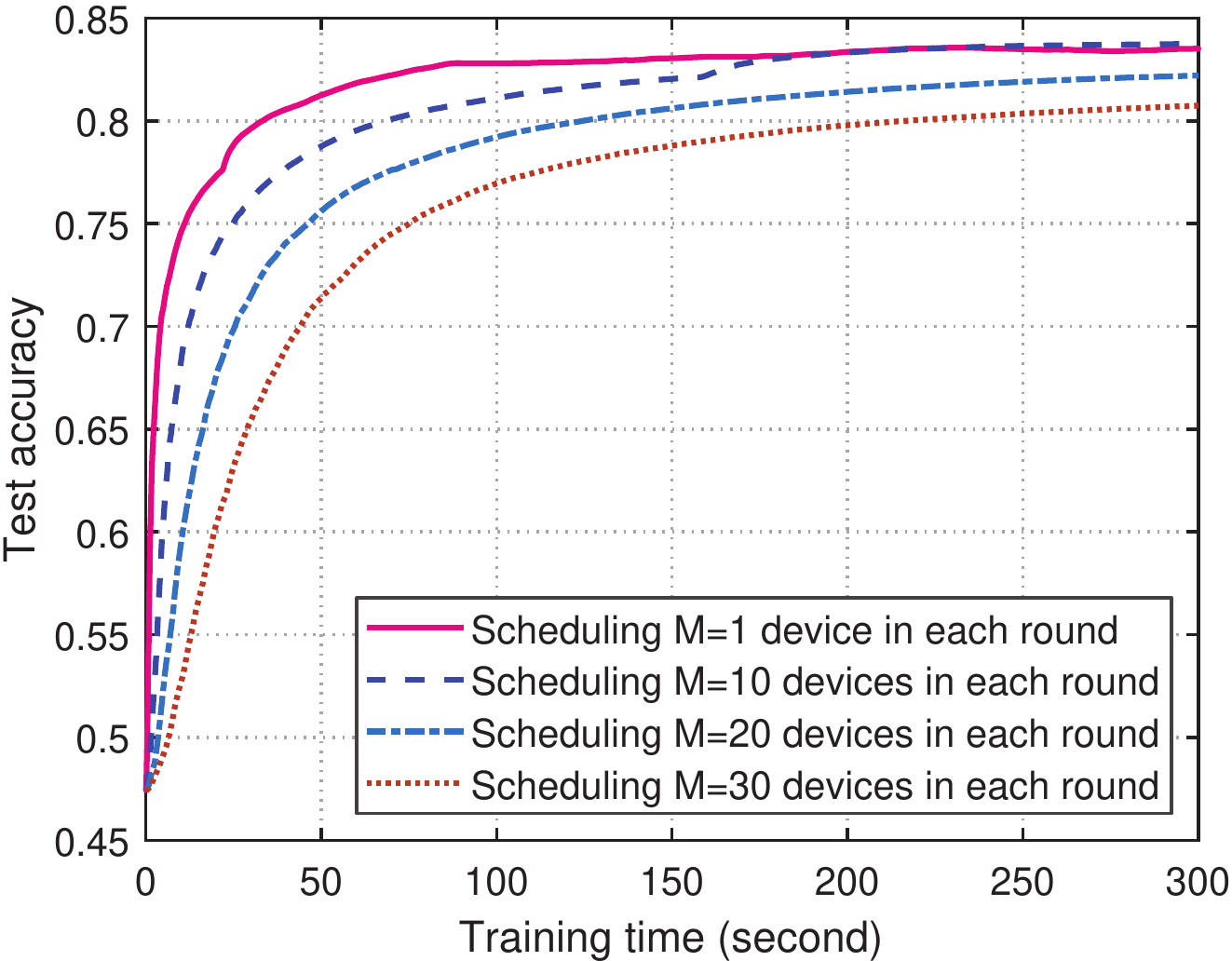}
        \label{Comparison 1MHz}
	}
	\subfigure[$B=20$ MHz]{
		\includegraphics[width=3.7in]{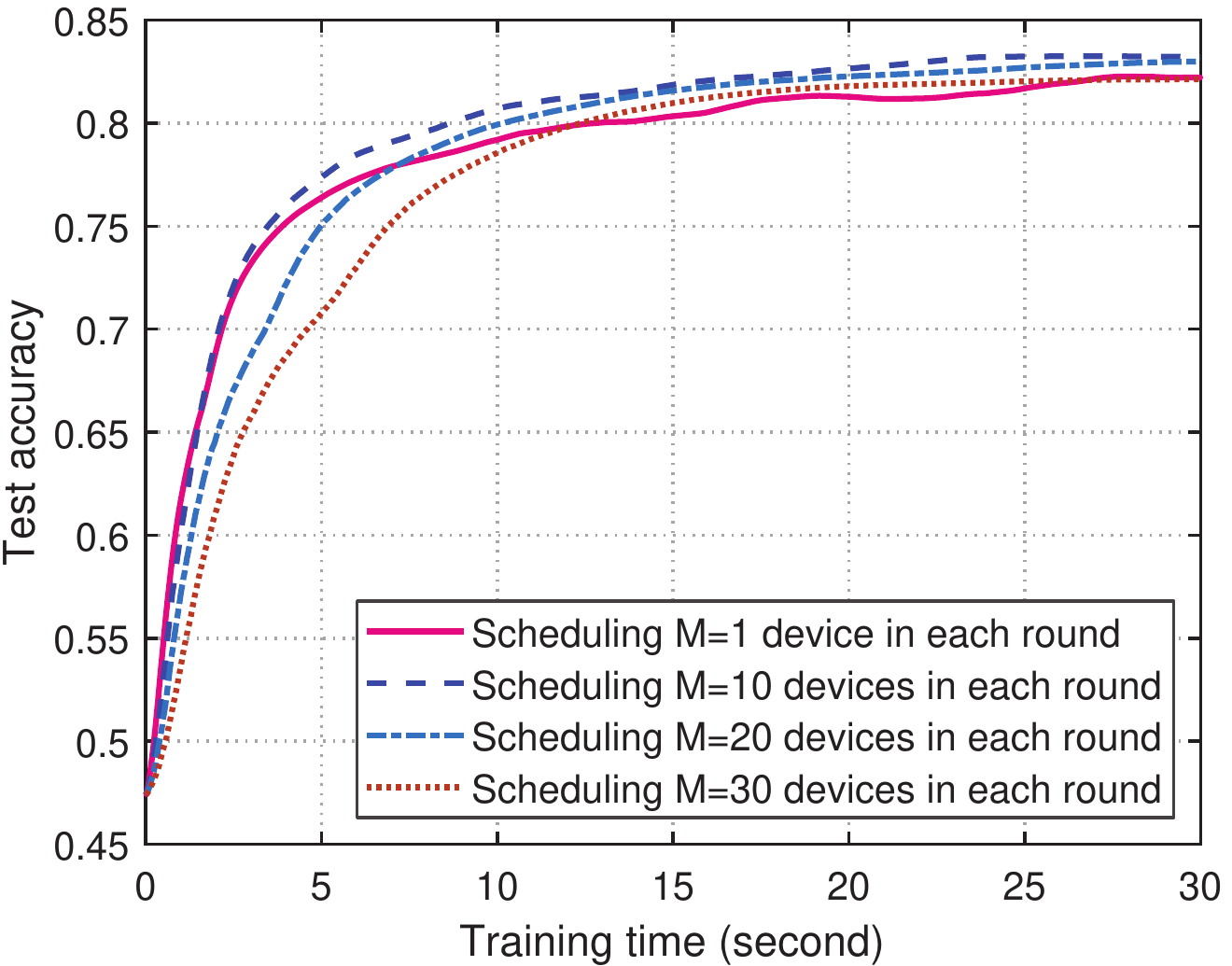}
        \label{Comparison 20MHz}
	}
	\caption{Performance comparison between two strategies.}
\end{figure}
\textit{2) Scheduled Device Number:} Given the alternative strategies of scheduling one and multiple devices in each communication round, a relevant concern is the proper selection between them. For exposition, we take the SVM model as an example and analyze the impact of the communication resource on model convergence. The curves of learning accuracy versus the training time with different scheduled device numbers are presented in Figs. \ref{Comparison 1MHz} and \ref{Comparison 20MHz}, for small bandwidth $B=1$ MHz and large bandwidth $B=20$ MHz, respectively. In Fig. \ref{Comparison 1MHz}, we can observe that the scheme of scheduling one device in each round can achieve faster convergence speed than other schemes without sacrificing the learning accuracy. In contrast, the scheme of scheduling 10 devices in each round is superior to other schemes in terms of convergence speed when the system bandwidth is $20$ MHz. This result has a profound and refreshing implication that the device number selection criterion should adapt to the wireless environment. Upon a certain guarantee on communication resource, scheduling multiple devices in each round can reduce the global gradient variance and thus accelerate the training process. This is also aligned with the previous discussions that the communication efficiency is the primary concern in current FEEL systems.

\section{Conclusion}
In this paper, we have proposed a new importance- and channel-aware scheduling policy for update aggregation in FEEL. We have developed a probabilistic scheduling framework to guarantee the unbiasedness of the aggregated global gradient as well as to speed up convergence. An optimal scheduling policy has been designed to incorporate the channel state and update importance for decision making. We have also provided the concrete convergence analysis. Moreover, the probabilistic scheduling framework has been extended to the strategy of scheduling multiple devices in each communication round and some practical implementation issues have been highlighted as well. Comprehensive experiments using real datasets substantiate the performance gain of the proposed scheduling policy as compared with the benchmark policies.

%\textcolor[rgb]{0.00,0.00,1.00}{At a high level, the current work contributes a new principle of exploiting update importance and channel condition for scheduling design in FEEL. It also opens several directions for further investigation, e.g., extension to non-convex loss functions. Moreover, some interesting issues, such as adaptive balance coefficient selection and importance- and channel-aware scheduling based on model-averaging also deserve future investigation.}

%At a higher level, the current work contributes a new principle of exploiting update importance for learning performance improvement in FEEL. It also opens several directions for further investigation. One direction is to speed up the learning convergence by dynamically adjusting the number of scheduled devices in the training duration. Another interesting direction is to integrate the importance-aware design principle with the lazy updating approach for achieving an energy-efficient edge learning system. Last but not least, the scheduling design in heterogeneous networks with various hardware variability and network connectivity is also a promising topic to be addressed.
\begin{appendices}
\section{Proof of Lemma 2}
First, we define an auxiliary function as
\begin{equation}\label{auxiliary function}
  G\left({\bf{w}}\right)= \frac{\ell}{2} {\bf{w}}^{\top}{\bf{w}} - L\left({\bf{w}}\right).
\end{equation}
By taking the second-order partial derivative of $G\left({\bf{w}}\right)$ on ${\bf{w}}$, we have
\begin{align}\label{transform}
  \frac{\partial^2 G\left({\bf{w}}\right)}{\partial{\bf{w}}^2} &= \ell - \frac{\partial^2 L({\bf{w}})}{\partial{\bf{w}}^2} \overset{(a)}\geq 0,
\end{align}
where ($a$) is due to (\ref{smooth assumption}). Therefore, $G\left({\bf{w}}\right)$ is convex on ${\bf{w}}$, leading to the following sufficient condition as
\begin{equation}\label{smooth apply}
  G\left({\bf{u}}\right) \geq  G\left({\bf{v}}\right) + \nabla G\left({\bf{v}}\right)^{\top} \left({\bf{u}}-{\bf{v}}\right),
\end{equation}
which is equivalent to
\begin{equation}\label{smooth apply2}
L\left(\bf{u}\right) \leq L\left(\bf{v}\right) + \nabla L\left(\bf{v}\right)^\top \left( \bf{u} - \bf{v}\right) + \frac{\ell}{2} \left\Vert \bf{u} - \bf{v} \right\Vert^2.
\end{equation}
On the other hand, the global model update at the edge server follows
\begin{equation}
	  {\bf{w}}^{t+1} = {\bf{w}}^{t} - \eta^t {\widehat{\bf{g}}^t}.
\end{equation}
Let ${\bf{u}}={\bf{w}}^{t+1}$ and ${\bf{v}}={\bf{w}}^t$ such that ${\bf{u}}-{\bf{v}} = - \eta^t {\widehat{\bf{g}}^t}$. Applying it into (\ref{smooth apply2}), we have
\begin{equation}\label{smooth apply3}
L\left({\bf{w}}^{t+1}\right) \leq L\left({\bf{w}}^{t}\right) + \left( {\bf{g}}^t \right)^\top \left(-\eta^t \widehat{\bf{g}}^t \right) +\frac{\ell}{2} \left\Vert -\eta^t \widehat{\bf{g}}^t \right\Vert^2.
	\end{equation}
Taking expectation in both sides of (\ref{smooth apply3}), it follows that
\begin{align}
	\mathbb{E}\left\{L\left({\bf{w}}^{t+1}\right)\right\} &\leq \mathbb{E}\left\{L\left({\bf{w}}^{t}\right)\right\} - \eta^t \left( {\bf{g}}^t \right)^\top \mathbb{E} \left\{\widehat{\bf{g}}^t \right\} +\frac{\ell}{2} \left( \eta^t \right)^2 \mathbb{E}\left\{ \left\Vert \widehat{\bf{g}}^t \right\Vert^2 \right\} \notag\\
	&=\mathbb{E}\left\{L\left({\bf{w}}^{t}\right)\right\} - \eta^t \left( {\bf{g}}^t \right)^\top \mathbb{E} \left\{\widehat{\bf{g}}^t \right\} +\frac{\ell}{2} \left( \eta^t \right)^2 \left[ \left(\mathbb{E}\left\{ \widehat{\bf{g}}^t  \right\} \right)^2 + \mathbb{V} \left\{ \widehat{\bf{g}}^t \right\}\right] \notag\\
	&\overset{(b)}{=} \mathbb{E}\left\{ L\left({\bf{w}}^t\right)\right\} - \eta^t \left( 1 - \frac{\ell}{2}  \eta^t \right) \left\Vert {\bf{g}}^t \right\Vert^2 + \frac{1}{2} \ell \left(\eta^t\right)^2 \mathbb{V}\left\{\widehat{\bf{g}}^t\}, \right. \label{One-round-proof}
\end{align}
where $ \mathbb{V} \left\{\widehat{\bf{g}}^t\right\} = \mathbb{E}\left\{ \left \Vert\widehat{\bf{g}}^t - {\bf{g}}^t\right \Vert^2\right\}$ is the variance of the global gradient and ($b$) is because $\widehat{\bf{g}}^t$ is an unbiased estimate of the ground-truth global gradient. Then subtracting $\mathbb{E}\left\{L\left({\bf{w}}^*\right)\right\}$ in both sides of (\ref{One-round-proof}), we can obtain the one-round convergence rate in Lemma 2.

\section{Proof of Lemma 3}
The objective function (\ref{P3a}) is convex because $\frac{1}{x}$ and $x$ are both convex on $\left(0, +\infty\right)$. Moreover, the constraints (\ref{P3b}) and (\ref{P3c}) are linear. Thus, $\mathscr{P}_3$ is convex and can be solved by the Karush–Kuhn–Tucker (KKT) conditions \cite{KKT}. The partial Lagrange function is defined as
\begin{equation}\label{Lagrangian funtion}
  \mathscr{L}= \sum_{k=1}^K a_k \frac{1}{x_k} +b_k x_k + \lambda \left(\sum_{k=1}^{K} c_k x_k -d\right),
\end{equation}
where $\lambda \geq 0$ is the Lagrange multiplier associated with the constraint (\ref{P3b}). Let $\left\{x_1^*, x_2^*, \cdots, x_K^*\right\}$ denote the optimal solution to $\mathscr{P}_3$. Then applying KKT conditions leads to the following necessary and sufficient conditions
\begin{align}
	 &\frac{\partial \mathscr{L}}{\partial x_k^*} = -\frac{a_k}{\left(x_k^*\right)^2} + b_k + \lambda^*c_k =
	\left\{
	\begin{array}{ll}
	\geq 0,~x_k^*=0\\
	=0,~x_k^*>0
	\end{array},~~~\forall k,
	\right.\\
	 &\sum_{k=1}^K c_k x_k^* - d = 0.
	\end{align}
By solving these equations, we can obtain the solution in Lemma 3.
\section{Proof of Theorem 2}
To prove Theorem 2, we first establish the following lemma.
\begin{lem}
The squared norm of the ground-truth global gradient in the $t$-th communication round satisfies
\begin{equation}
	\left\Vert {\bf{g}}^{t} \right\Vert^2 \geq 2\mu \left( L\left({\bf{w}}^{t}\right) -L\left({\bf{w}}^*\right) \right). \label{inequality}
\end{equation}
\end{lem}
\vspace{-0.4em}

\proof
Since the global loss function is strongly convex with the parameter $\mu$, it follows
\begin{align}
L\left({\bf{w}}^{t+1}\right) \geq L\left({\bf{w}}^{t}\right) + \left({\bf{g}}^t\right)^\top \left( {\bf{w}}^{t+1} - {\bf{w}}^{t}  \right) +\frac{\mu}{2} \left\Vert {\bf{w}}^{t+1} - {\bf{w}}^{t} \right\Vert^2.  \label{strong_convex}
\end{align}
Minimizing both sides of (\ref{strong_convex}) with respect to ${\bf{w}}^{t+1}$, we have
\begin{align}
\min_{{\bf{w}}^{t+1}} L\left({\bf{w}}^{t+1}\right) \geq \min_{{\bf{w}}^{t+1}} \left[L\left({\bf{w}}^{t}\right)+ \left({\bf{g}}^t\right)^\top \left( {\bf{w}}^{t+1} - {\bf{w}}^{t}  \right) +\frac{\mu}{2} \left\Vert {\bf{w}}^{t+1} - {\bf{w}}^{t} \right\Vert^2 \right].  \label{strong_convex_2}
\end{align}
The minimization of the left-hand side of (\ref{strong_convex_2}) is achieved when ${\bf{w}}^{t+1}={\bf{w}}^*$ while the right-hand side of (\ref{strong_convex_2}) is minimized when ${\bf{w}}^{t+1} = {\bf{w}}^t - \frac{1}{\mu} {\bf{g}}^t$ \cite{Minimizer}. Thus, it leads to
\begin{align}
L\left({\bf{w}}^*\right) \geq L\left({\bf{w}}^{t}\right) - \frac{1}{2\mu} \left\Vert {\bf{g}}^{t} \right\Vert^2, \label{strong_convex_3}
\end{align}
which is equivalent to the desired result in (\ref{inequality}).
%\end{IEEEproof}

Based on Lemma 5, we can prove Theorem 2 as follows. Combining (\ref{inequality}) with the one-round convergence rate in     (\ref{One-round-proof}), we can derive that
\begin{align}
&\mathbb{E}\left\{ L\left({\bf{w}}^{t+1}\right)-L\left(\bf{w^*}\right)\right\} \notag\\
&\leq
\mathbb{E}\left\{L\left({\bf{w}}^{t}\right) -L\left(\bf{w^*}\right)\right\} - 2\mu \eta^t \mathbb{E}\left\{\left( L\left({\bf{w}}^{t}\right) -L\left({\bf{w}}^*\right) \right)\right\} +\frac{\ell}{2} \left( \eta^t \right)^2 \mathbb{E}\left\{ \left\Vert \widehat{\bf{g}}^t \right\Vert^2 \right\}\notag\\
&=
\left(1-2\mu \eta^t\right) \mathbb{E}\left\{ L\left({\bf{w}}^t\right)-L\left(\bf{w^*}\right)\right\} +\frac{\ell}{2} \left( \eta^t \right)^2 \mathbb{E}\left\{ \left\Vert \widehat{\bf{g}}^t \right\Vert^2 \right\}\notag\\
&\overset{\text{recursively}}{\leq} \prod_{i=1}^t \left(1-2\mu \eta^i\right) \mathbb{E}\left\{ L\left({\bf{w}}^1\right)-L\left(\bf{w^*}\right)\right\}  +  \frac{\ell}{2} \sum_{i=1}^t A^i \left( \eta^i\right)^2 \mathbb{E}\left\{ \left\Vert \widehat{\bf{g}}^i \right\Vert^2 \right\}, \label{convergece rate in corolary}
\end{align}
where $A^i=\prod_{j=i+1}^{t} \left(1-2\mu \eta^j\right)$. Then applying the optimal scheduling policy into $\mathbb{E}\left\{ \left\Vert \widehat{\bf{g}}^i \right\Vert^2 \right\}$, we can obtain the desired result in Theorem 2, which ends the proof.	

\section{Proof of Corollary 1}
We will prove Corollary 1 by induction. First, according to (\ref{convergece rate in corolary}), the following inequality holds in each communication round, as
\begin{equation}
\mathbb{E}\left\{ L\left({\bf{w}}^{t+1}\right)-L\left(\bf{w^*}\right)\right\} \leq \left(1-2\mu \eta^t\right) \mathbb{E}\left\{ L\left({\bf{w}}^t\right)-L\left(\bf{w^*}\right)\right\} +\frac{\ell}{2} \left( \eta^t \right)^2 \mathbb{E}\left\{ \left\Vert \widehat{\bf{g}}^t \right\Vert^2 \right\},
\end{equation}
For the diminishing learning rate $\eta^t = \dfrac{\chi}{t+\nu}$ where $\chi > \dfrac{1}{2 \mu}$ and $\nu > 0$, we have
\begin{itemize}
  \item For $t=1$, it satisfies
\begin{align}
\dfrac{\zeta}{t+\nu}\Big|_{t=1} &= \dfrac{\zeta}{1+\nu} \notag\\
&\overset{(c)}{\geq} \dfrac{\left(1+\nu\right) \mathbb{E}\left\{L\left({\bf{w}}^1\right)-L\left(\bf{w^*}\right)\right\}}{1+\nu} \notag\\
&= \mathbb{E}\left\{L\left({\bf{w}}^1\right)-L\left(\bf{w^*}\right)\right\},
\end{align}
where ($c$) is derived from the definition of $\zeta$. Therefore, the desired result holds when $t=1$.

\item Assume that the desired result holds for some $t = N \geq 1$, i.e., $\mathbb{E}\left\{ L\left({\bf{w}}^N\right)-L\left(\bf{w^*}\right)\right\} \leq \dfrac{\zeta}{N+\nu}$. Define $G = \max \left\Vert \widehat{\bf{g}}^t \right\Vert $ as the upper bound of the global gradient norm. Then when $t=N+1$, it follows
\begin{align}
\mathbb{E}&\left\{ L\left({\bf{w}}^{N+1}\right)-L\left(\bf{w^*}\right)\right\} \notag
\\&\leq \left(1-2\mu \eta^N\right) \mathbb{E}\left\{ L\left({\bf{w}}^N\right)-L\left(\bf{w^*}\right)\right\} +\frac{\ell}{2} \left( \eta^N \right)^2 \mathbb{E}\left\{ \left\Vert \widehat{\bf{g}}^N \right\Vert^2 \right\}\notag\\
& \leq \left(1-2\mu \dfrac{\chi}{N+\nu} \right) \dfrac{\zeta}{N+\nu} + \frac{\ell}{2} \left(\dfrac{\chi}{N+\nu} \right)^2 G^2\notag\\
& = \dfrac{\zeta \left(N+\nu-1\right) + \dfrac{1}{2} \ell G^2 \chi^2 - \left(2\mu\chi-1\right)\zeta}{\left(N+\nu\right)^2}\notag\\
& \leq \dfrac{\zeta \left(N+\nu-1\right) + \dfrac{1}{2} \ell G^2 \chi^2 - \left(2\mu\chi-1\right)\dfrac{\ell G^2 \chi^2}{2\left(2 \mu \chi -1 \right)}}{\left(N+\nu\right)^2}\notag\\
& = \dfrac{\zeta \left(N+\nu-1\right)}{\left(N+\nu\right)^2}\notag\\
& \leq \dfrac{\zeta}{N+1+\nu}.
\end{align}
Hence, the desired result holds for $t=N+1$ if it holds for $t=N$. Towards this end, we can conclude that $\mathbb{E}\left\{ L\left({\bf{w}}^T\right)-L\left(\bf{w^*}\right)\right\} \leq \dfrac{\zeta}{T+\nu}$ holds for $\forall T \in \mathbb{Z}^+$, which completes the proof.
\end{itemize}
\end{appendices}

\end{document}